\documentclass[5p, two column, authoryear, letter, 10pt]{elsarticle}

\usepackage{mathtools}
\usepackage{amsfonts}
\usepackage{lineno}

\modulolinenumbers[5]

\usepackage{subfigure}
\usepackage{amssymb}
\usepackage{tikz}

\usepackage{array,booktabs}
\newcolumntype{x}[1]{>{\centering\arraybackslash\hspace{0pt}}p{#1}}
\newcolumntype{y}[1]{>{\hfill}p{#1}}

\journal{Annals of Nuclear Energy}

\DeclarePairedDelimiter\abs{\lvert}{\rvert}%
\usepackage{numcompress}\biboptions{authoryear}

\usepackage{xspace} 

\newcommand{\TLE}{TL estimator\xspace}  
\newcommand{\TLEPERIOD}{TL estimator\@}  

\newcommand{\EVE}{EV estimator\xspace}  
\newcommand{\EVEPERIOD}{EV estimator\@}  

\newcommand{\VRCE}{VRC estimator\xspace} 
\newcommand{\VRCEPERIOD}{VRC estimator\@} 

\newcommand{\MCNP}{MCNP6\textsuperscript{\textregistered}\xspace}
\newcommand{\MCNPPERIOD}{MCNP6\textsuperscript{\textregistered}\@}

\newcommand{\Nvidia}{NVIDIA\textsuperscript{\textregistered}\xspace}

\def \keff {$k_\text{eff}$}

\newcommand{\CPP}
{C\nolinebreak[4]\hspace{-.05em}\raisebox{.22ex}{\footnotesize\bf ++}}

\usepackage[colorlinks]{hyperref} 

\AtBeginDocument{%
\hypersetup{
            colorlinks = true,
            linkcolor = cyan,
            urlcolor  = cyan,
            citecolor = cyan,
            anchorcolor = cyan}}

\def\equationautorefname~#1\null{Eq. (#1)\null}

\let\oldhref\href
\renewcommand{\href}[2]{\oldhref{#1}{\hbox{#2}}}

\newcommand\solidrule[1][0.5cm]{\rule[0.5ex]{#1}{.4pt}}
\newcommand\dashedrule{\mbox{%
  \solidrule[1.5mm]\hspace{0.8mm}\solidrule[0.4mm]\hspace{0.8mm}\solidrule[1.5mm]}}

\begin{document}

\begin{frontmatter}

\title{A Monte Carlo Volumetric-Ray-Casting Estimator for Global Fluence Tallies on GPUs\tnoteref{notice1} }
\tnotetext[notice1]{This work has been authored by an employee of Los Alamos National Security, LLC, operator of the Los Alamos National Laboratory under Contract No. DE-AC52-06NA25396 with the U.S. Department of Energy.  The United States Government retains and the publisher, by accepting this work for publication, acknowledges that the United States Government retains a nonexclusive, paid-up, irrevocable, world-wide license to publish or reproduce this work, or allow others to do so for United States Government purposes.}

\author{Jeremy E. Sweezy}
\ead{jsweezy@lanl.gov}
\address{Los Alamos National Laboratory, P.O. Box 1663, MS A143, Los Alamos, NM 87545}

\begin{abstract}
A Monte Carlo fluence estimator has been designed to take advantage of the computational power of  graphical processing units (GPUs)\@.   
This new estimator, termed the volumetric-ray-casting estimator, is an extension of the expected-value estimator.  
It can be used as a replacement of the track-length estimator for the estimation of global fluence.
Calculations for this estimator are performed on the GPU while the Monte Carlo random walk is performed on the central processing unit (CPU)\@.  
This method lowers the implementation cost for GPU acceleration of existing Monte Carlo particle transport codes as there is little modification of the particle history logic flow\@.    
Three test problems have been evaluated to assess the performance of the volumetric-ray-casting estimator for neutron transport on GPU hardware in comparison to the standard track-length estimator on CPU hardware\@.  
%
%
Evaluation of neutron transport through air in a criticality accident scenario showed that the volumetric-ray-casting estimator achieved 23 times the performance of the track-length estimator using a single core CPU paired with a GPU and 15 times the performance of the track-length estimator using an eight core CPU paired with a GPU\@.
Simulation of a pressurized water reactor fuel assembly showed that the performance improvement was 6 times within the fuel and 7 times within the control rods using an eight core CPU paired with a single GPU\@.

 
 \end{abstract}


\end{frontmatter}




\section{Introduction}

Accelerators augmenting central processing units (CPUs) is currently one of the possible paths towards exascale computing.  
Sixty one of the 500 fastest supercomputers in the world, including ORNL's Titan (the 3rd fastest), use GPU accelerators~\citep{top500b}.  
Several research groups have demonstrated specially built Monte Carlo particle transport on GPUs~\citep{gpumcd, henderson2013, bergmann2015,  ARCHER} with significant increases in performance.  
However, a general Monte Carlo code,  such as LANL's \MCNP\citep{mcnp6}  code, represents hundreds of person-years of development and porting the entire code base to support specialized hardware is not feasible.  

\begin{figure}[bt]\centering 
\includegraphics[width=\linewidth,trim=0.0cm 0.0cm 0.0cm 0.0cm,clip]{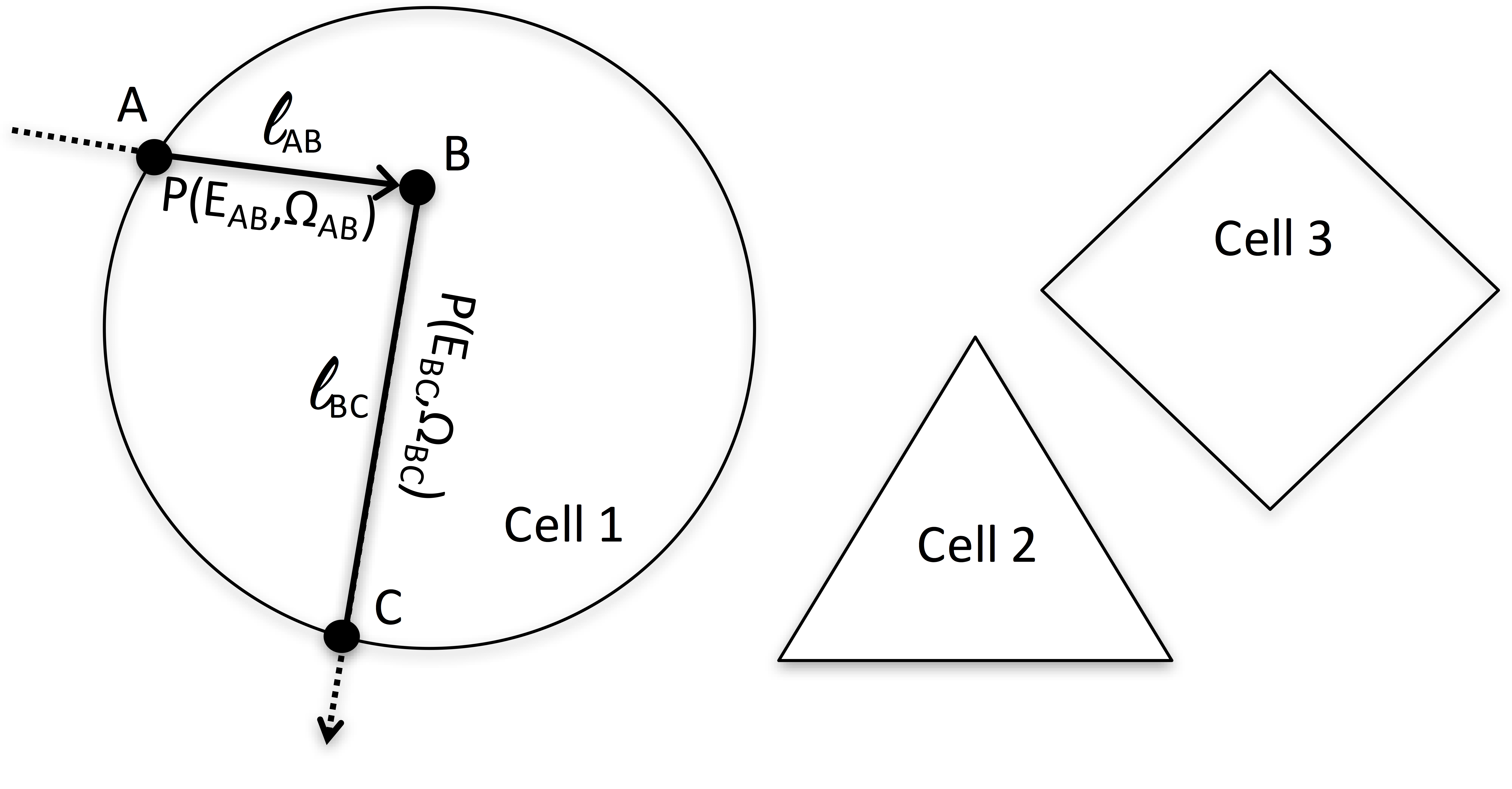}
\caption{Track-length estimator.  A random walk particle enters Cell 1 at point A, has a collision at point B, and exits Cell 1 at Point C.  The \TLE makes two contributions along the random walk, one at energy $E_{AB}$ with length $l_{AB}$ and one at energy $E_{BC}$ with length $l_{BC}$. }
\label{fig:track_length}
\end{figure}

Traditionally, Monte Carlo methods have been used to calculate a few local fluence values. Today, the challenge for Monte Carlo codes is to calculate global particle fluence~\citep{martin2012}.  
Global fluence rates are needed to calculate local power densities for nuclear reactor core design, dose rates in facilities, neutron activation of reactor components, treatment planning for radiotherapy,  and many other applications.    The track-length (TL) estimator has been used as the standard estimator for calculating global fluence rates since the mid-1960's~\citep{gelbard}.   During the random walk, the particle's path length in each cell contributes to the track-length estimate of fluence (\autoref{fig:track_length}).  The \TLE has the advantage that no additional values must be calculated, the score for a cell is the distance a particle travels in the cell. 

\begin{figure}[tb]\centering 
\includegraphics[width=\linewidth]{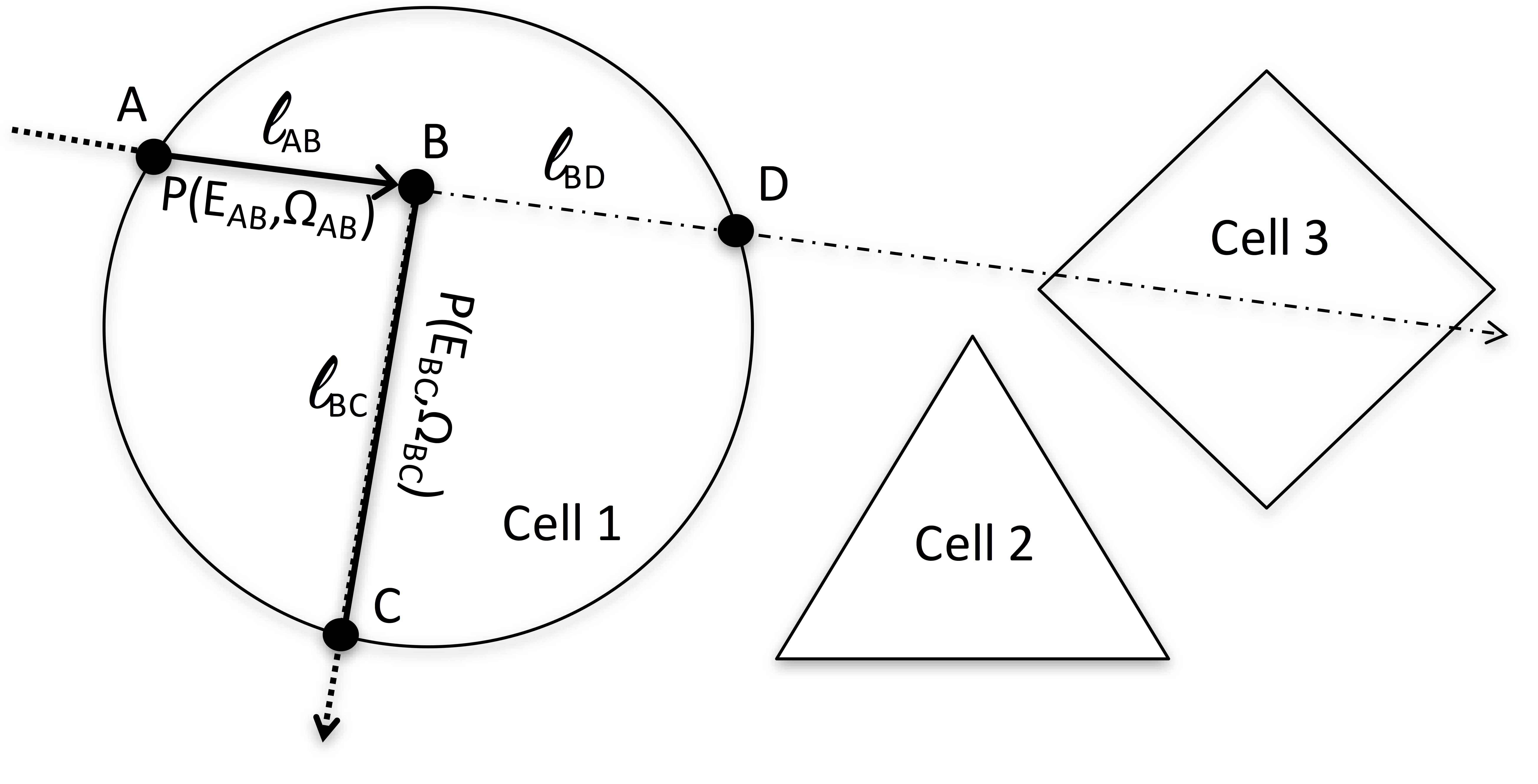}
\caption{Expected-value estimator.  A random walk particle enters Cell 1 at point A, has a collision at point B, and exits Cell 1 at Point C\@.  
The expected-value estimator makes two contributions to cell 1 using Eq.~\eqref{eq:expected_value}, one contribution at energy $E_{AB}$ with length $l_{AD} = l_{AB} + l_{BD}$ and another contribution at energy $E_{BC}$ and with length $l_{BC}$.}
\label{fig:expected_value}
\end{figure}

\cite{spanier:66} points out that \TLE is an unbiased estimator of the expected path length of a particle's flight through a cell.  The expected path length can be calculated directly:    
\begin{equation}
F(i,E) = \frac{W} {\Sigma_{t,i}(E)} \left[ 1 - \exp{ \left(-\Sigma_{t,i}(E) l_i \right)} \right], 
\label{eq:expected_value}
\end{equation}
where $i$ is the tally cell, $E$ is the energy of the particle after collision, $W$ is the statistical weight of the particle, $\Sigma_{t,i}(E)$ is the total cross-section of cell $i$ at energy $E$, and $l_i$ is the ray length (distance from entrance to possible exit) through cell $i$\@.    
Eq. \eqref{eq:expected_value} is the  expected-value (EV) estimator (\autoref{fig:expected_value}).
It is estimator XI in \cite{gelbard}\@.   
\cite{macmillan} found that the \EVE generally provided a lower variance than the \TLE, but considered it too expensive due to the cost of calculating the exponential function.   
Macmillian wrote:
\begin{quote}
\textit{The estimator, in spite of its excellent performance where scattering is small, is not attractive for general use, because, where scattering is large, it combines mediocre performance with relatively large computing time.}
\end{quote}
The state of computing has changed since Macmillian's paper in 1966\@.   
New computing hardware provides opportunities to reconsider computational techniques that were once considered too expensive.    

\section{ Volumetric-Ray-Casting Estimator }

The \EVE can be extended to contribute to cells in which the random walk does not pass through.   
This is evident in \autoref{fig:expected_value}, the ray along A-B also intersects Cell 3\@.   
The contribution to Cell 3 can also be calculated from Eq.~\eqref{eq:expected_value}, but the statistical weight must be modified to account for attenuation of the ray prior to entering the cell\@.   
The statistical weight entering cell $i$, $W_i(E)$, is calculated from the optical thickness between the collision or source point and the point the ray enters the cell:

\begin{equation}
W_{i}(E) = W \exp{ \left[ - \int_{0}^{\abs{ \boldsymbol{r'} - \boldsymbol{r}}} \Sigma_{t} (\boldsymbol{r}+\boldsymbol{\hat{\Omega}} s,E) ds \right] } ,
\label{eq:expected_value_weight}
\end{equation}
where $W_{i}(E)$ is the statistical weight of the particle entering cell $i$, $i$ is the tally cell, $E$ is the energy of the particle emitted in direction $\boldsymbol{\hat{\Omega}}$, $W$ is the statistical weight of the particle at collision, $\boldsymbol{r}$ is the collision point, $\boldsymbol{r'}$ is the point that the ray enters cell $i$\@.  
The integral expression is the optical thickness. 
The weight, $W$, in Eq.~\eqref{eq:expected_value} is replaced with the weight entering cell $i$, Eq.  ~\eqref{eq:expected_value_weight}, resulting in:

\begin{equation} 
\begin{split}
F(i,E) = \frac{W \left[ 1 - \exp{ \left(-\Sigma_{t,i}(E) l_i \right)} \right] } {\Sigma_{t,i}(E)} \\
\times \exp{ \left[ - \int_{0}^{\abs{ \boldsymbol{r'} - \boldsymbol{r}}} \Sigma_{t} (\boldsymbol{r}+\boldsymbol{\hat{\Omega}} s,E) ds \right] } . 
\end{split}
\label{eq:ray_trace_1}
\end{equation} 
The unique property of the \EVE expressed in Eq.~\eqref{eq:ray_trace_1} is that it scores to cells other than the cells along the path of the random walk.
The use of this type of \EVE as an expected leakage estimator is quite old~\citep{KSCHWENDT}.
More recently it has been used to improve estimates of fluence and reaction rates in cells that see few random walk particles~\citep{Mosher:2010}.  
A modification of the \EVE, Eq.~\eqref{eq:ray_trace_1}, will be shown to be an alternative to the \TLE for estimating global fluence rates. 
The type of ray casting through volumes described by Eq.~\eqref{eq:ray_trace_1} is a standard technique for physically based image rendering~\citep{Krivanek2014}.    

\begin{figure}[tb]\centering 
\includegraphics[width=\linewidth]{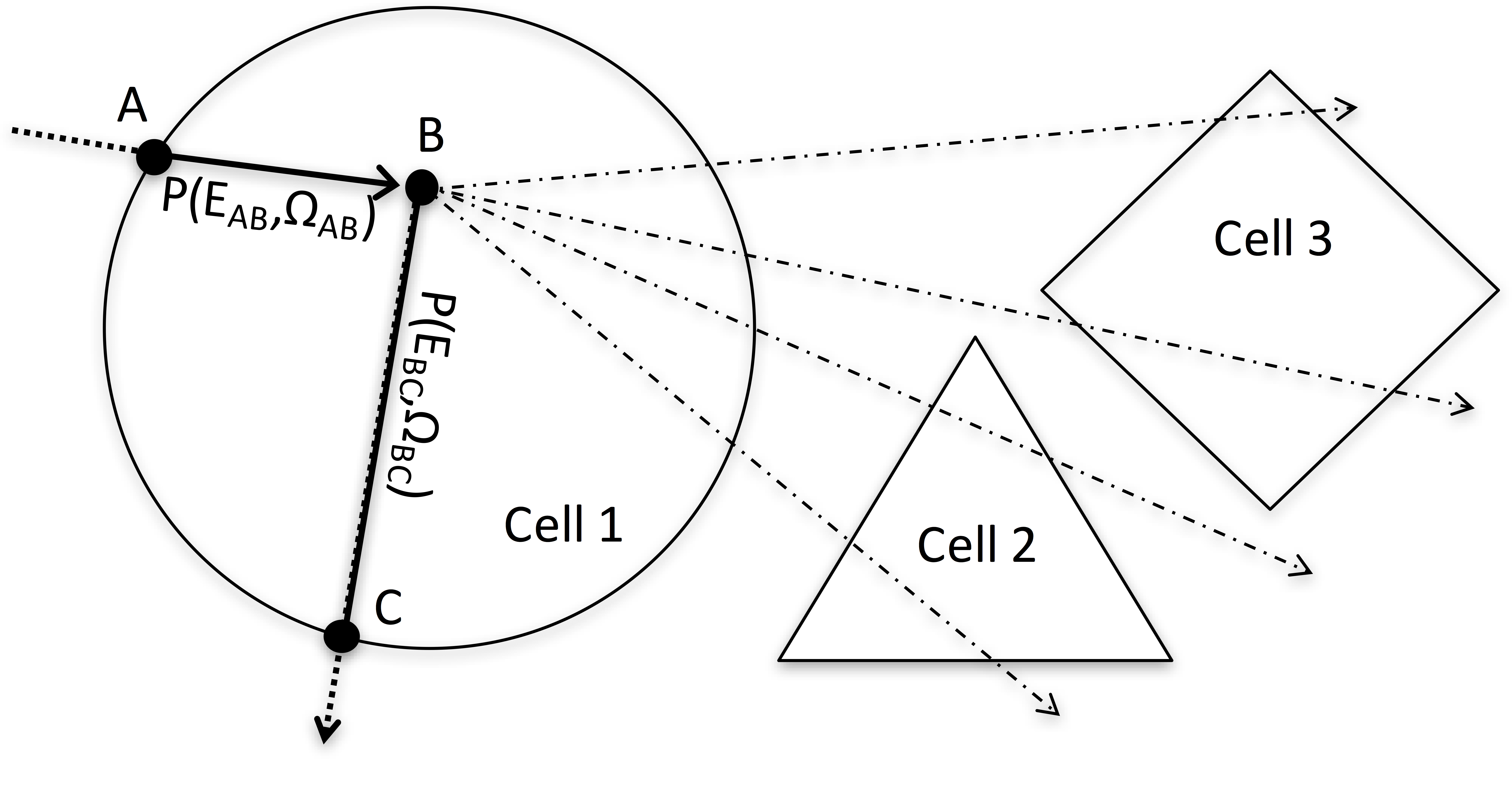}
\caption{\VRCE.   Multiple out-going rays can be sampled from each collision event and each ray can score to multiple cells.}
\label{fig:ray_trace_estimator}
\end{figure}

With the advent of GPUs, we can consider not just a single ray per collision but instead sample many possible rays (\autoref{fig:ray_trace_estimator}).   
If multiple rays per source or collision event are sampled, then the statistical weight of each ray must be reduced appropriately.   This results in an estimate for each ray, $j$:
\begin{equation}
\begin{split}
F_{j}(i,E_{j}) = \frac{W \left[ 1 - \exp{ \left(-\Sigma_{t,i}(E_{j}) l_{i,j} \right)} \right] } {N ~ \Sigma_{t,i}(E_{j})} 
\\
\times \exp{ \left[ - \int_{0}^{\abs{ \boldsymbol{r'} - \boldsymbol{r}}} \Sigma_{t} (\boldsymbol{r}+\boldsymbol{\hat{\Omega}_{j}} s',E_{j}) ds' \right] } 
\\ 
\mbox{for} \;\;\;  j=1, 2, ...,N ~~,
\end{split}
\label{eq:ray_trace_2}
\end{equation}
where $N$ is the number of out-going rays sampled per collision, $E_j$ is the energy of ray $j$, $\boldsymbol{\hat{\Omega}_{j}}$ is the unit direction of ray $j$, and $l_{i,j}$ is the length of ray $j$ through cell $i$.  
The estimator described by Eq.~\eqref{eq:ray_trace_2} will be referred to as the volumetric-ray-casting  (VRC) estimator to differentiate it from the \EVEPERIOD.   
However, it should be noted that the \VRCE is merely the \EVE used in the extreme.  

The power of the \VRCE comes from the fact that \textit{more information is obtained from each particle collision} than from the \TLEPERIOD.   
When using a traditional CPU this power is negated by the additional computational expense of performing the ray casting.   
The proposed method off-loads the ray casting to the GPU and hides any additional expense by performing the ray casting concurrent with the random walk on the CPU\@.     

\section{Implementation of the \VRCE}

A library, called MonteRay, has been written using CUDA C/\CPP\ \citep{cuda} to evaluate the \VRCE on GPU hardware.  
MonteRay uses continuous energy total cross section data and can transport particles in 3-D Cartesian mesh geometry.
MonteRay has been coupled with Los Alamos National Laboratory's newest Monte Carlo particle transport code, the Monte Carlo Application Toolkit (MCATK), pronounced ``mac-attack''\citep{mcatk}\@.
MCATK, a modular \CPP\ code,  has been in development since 2008\@.    
It is capable of transporting neutrons and photons using continuous energy ACE formatted cross section data.  
    
\begin{figure}[tbph!]\centering 
\includegraphics[height=0.9\textheight]{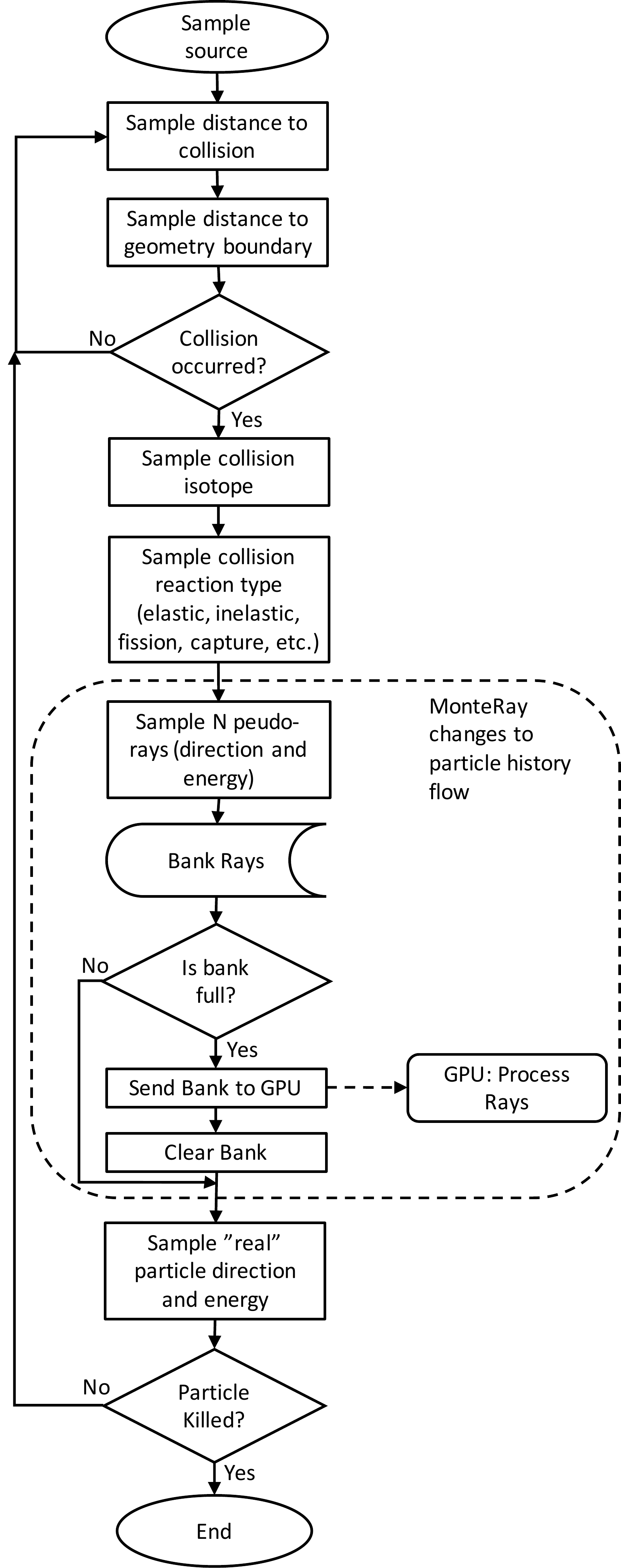}
\caption{Flow chart of changes to the Monte Carlo particle history flow for the addition of MonteRay for the calculation of the \VRCE. }
\label{fig:flow_chart}
\end{figure}

The random walk was performed with MCATK on the CPU and rays from collisions were sampled on the CPU\@.   
When sampling multiple rays per collision, the collision isotope and collision reaction channel was not resampled.   
Instead, the outgoing particle direction and particle energy from the chosen collision reaction channel was sampled multiple times. 
The rays were stored in a buffer and transferred to the GPU when the buffer became full.   
For assessing the performance of multiple CPU cores paired with a single GPU, MCATK was executed with MPI, and a single collision buffer was shared among the MPI processes of a node using MPI-3 shared memory.   
Only a single call to MonteRay, besides data transfer, was inserted into the Monte Carlo particle history loop(\autoref{fig:flow_chart}).

\begin{figure}[tb]\centering 
\includegraphics[height=0.578\textheight]{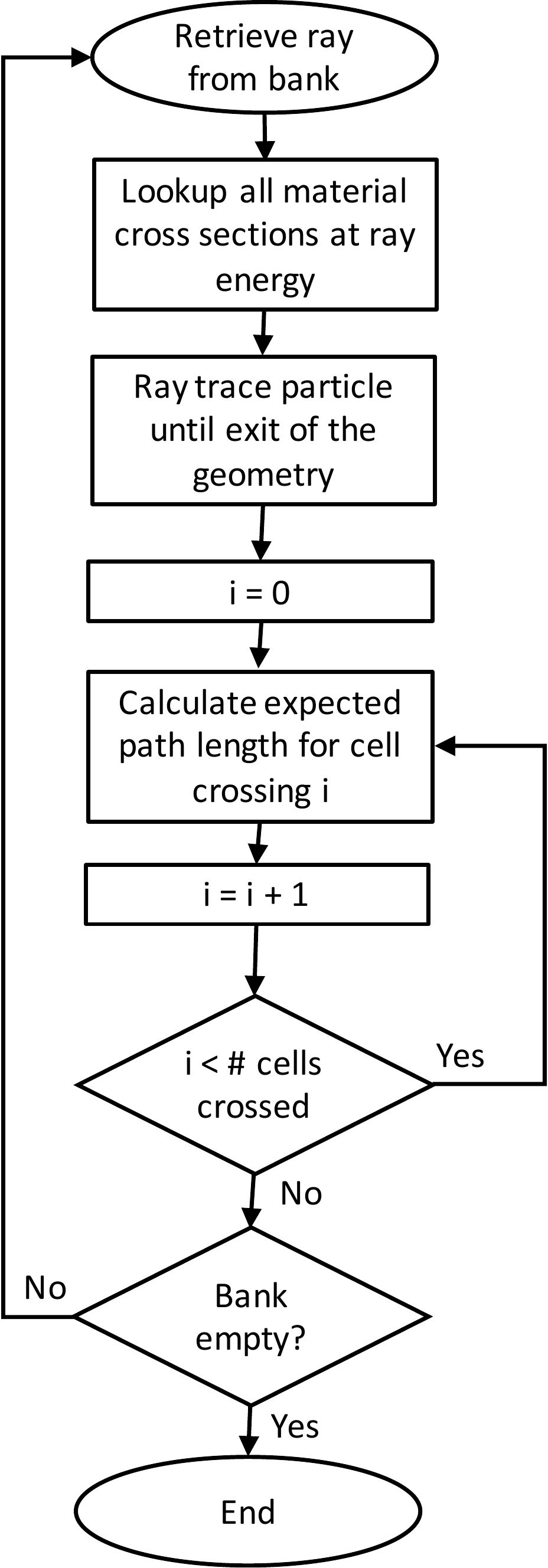}
\caption{Flow chart of GPU implementation of the \VRCE estimator. }
\label{fig:gpu_flow_chart}
\end{figure}

After the ray buffer was transferred to the GPU, the CPU cleared the buffer and resumed the simulation of additional particle histories.  
While the CPU was busy simulating more particles the GPU processed the rays stored in the buffer.  
Each GPU thread processed one individual ray at a time, first calculating all cross-sections, then ray casting the particle, and finally calculating the \VRCE (\autoref{fig:gpu_flow_chart})\@.  
The first step, cross-section lookup, was performed using the hash-based cross-section lookup scheme described by \cite{brownHash}.
The cross-section data was linearly interpolated between table values.  
ACE formatted cross-section data has been produced by NJOY to ensure that a linear interpolation reconstructs the data within a 0.1\% error~\citep{njoy}.
For the second step, each thread cast the ray through the entire geometry of the problem storing the path length travelled in each cell.  
In a final step, each GPU thread calculated the  \VRCE contribution to every cell crossed by the ray using its cross-section and ray path length from Eq.~\eqref{eq:ray_trace_2}. 
The tally remained on the GPU until the termination of the simulation, when it was transferred back to the CPU. 

The calculation of Eq.~\eqref{eq:ray_trace_2} was performed using double-precision operations to maintain accuracy, but using single precision cross-sections and distances.   
To maximize the speed of the calculations, the distances where calculated using single-precision ray casting operations.  
The neutron fluence rates calculated using single precision cross-sections and single precision ray casting on the GPU were compared with neutron fluence rates calculated using double precision cross-sections and double precision ray casting on the CPU for each test problems.   

\section{Evaluation Method}

The efficiency of Monte Carlo variance reduction techniques and tally methods are typically measured using a figure of merit (FOM) which accounts for both the computation time and the obtained tally variance.  
The FOM is defined as:
\begin{equation}
FOM = \frac{1} { \sigma^2 T} ,
\label{eq:fom}
\end{equation}
where $\sigma^2$ is the variance of the tally and T is the computational time (the standard deviation of the tally mean, $\sigma$, is the square root of the variance).   
A larger FOM indicates a more efficient method.    
To measure the performance of the \VRCE ($\eta$) the ratio of the \VRCE FOM ($\text{FOM}_{\text{VRCE}}$) to the \TLE FOM ($\text{FOM}_{\text{TLE}}$) was used.   
\begin{equation}
\eta =  \frac{ \text{FOM}_{\text{VRCE}} } {  \text{FOM}_{\text{TLE}} }  = 
\frac {\sigma_{\text{TLE}}^2  } { \sigma_{\text{VRCE}}^2 }
\frac {T_{\text{TLE}} }  {T_{\text{VRCE}} } 
\label{eq:performance}
\end{equation}

The time used in Eq.~\eqref{eq:performance} was the time used for the particle transport portion of the calculations.   
The problem setup and finalization portions were not included in the timings as they should be nearly equivalent for both methods.  
Additionally, the calculations were sufficiently long that the problem setup and finalization portions used negligible time compared to the particle transport portion.   
Timings were performed with the C \texttt{clock\_gettime} system call.  Timings ranged from a minimum of 23 seconds to a maximum of 3536 seconds. 
   
The performance of the \VRCE was evaluated for 1-20 CPU cores matched with a GPU\@.   
The \TLE FOM was measured with the same number of CPU cores.  
Two reference points will be used for the purposes of this study,  a single CPU core matched with a single GPU and eight CPU cores matched with a single GPU\@.   
The performance of the \VRCE for other numbers of CPU cores matched to a GPU have been included in tables and figures.  As a reference, the Titan supercomputer at ORNL currently uses 16 CPU cores per GPU~\citep{titanornl}\@.
It is expected that future GPU accelerated supercomputers will have 8-12 CPU cores per GPU\@.   

CPU timings were performed using 10 core Intel Haswell CPUs (E5-2660 v3) running at 2.60 GHz\@.   
There were 2 CPUs (20 total cores) per compute node.   
The GPU timings were performed using the same 2 Intel Haswell CPUs paired with an GeForce GTX TitanX GPU, which uses the \Nvidia Maxwell architecture.  
The TitanX GPU has 3072 CUDA cores operating at 1.0 GHz and has 12 GB of memory~\citep{titanx}\@. 
A NVIDIA Tesla K40 was also tested but was two times slower than the TitanX. 

In order to validate the use of single-precision ray casting for calculation of the \VRCE on the GPU, comparisons have been made to double-precision ray casting for \VRCE calculations on the CPU\@.  
Two metrics were used for this comparison.  The relative difference in the neutron fluence between ray casting in single-precision on the GPU and ray casting in double-precision on the CPU is , $\Delta_{\phi}(i)$:
\begin{equation}
\Delta_{\phi}(i) = \frac{ |\phi_{CPU}(i) - \phi_{GPU}(i)|} { \phi_{CPU}(i)}  ,
\label{eq:frac_diff}
\end{equation}
where $\phi_{CPU}(i)$ is the neutron fluence in cell $i$ calculated with double precision ray casting on the CPU, and $\phi_{GPU}(i)$ is the neutron fluence in cell $i$ calculated with single precision ray casting on the GPU\@.  The relative difference gives a measure of the difference that can be compared between cells of varying magnitudes.
The second metric is the fractional difference, $\epsilon_{\phi}(i)$, which is the ratio of the absolute difference to the standard deviation of the fluence:
\begin{equation}
\epsilon_{\phi}(i) =  \frac{ |\phi_{CPU}(i) - \phi_{GPU}(i)|} { \sigma(i) } ,
\label{eq:frac_diff_uncertainty}
\end{equation}
where $\sigma(i)$ is the standard deviation of the fluence in cell $i$.   
The fractional difference indicates if the difference is statistically significant.

\section{Tests}

Three test problems were considered:   
a 16x16 pressurized water reactor (PWR) fuel assembly, a simulation of a criticality accident in a concrete room, and a model of the reflected Godiva criticality benchmark problem.   
In all three cases, the neutron fluence was calculated with the VRC and TL estimators, and their relative performance was evaluated.    For each of the three tests, four analysis are presented:  the performance as a function of the number of rays sampled per collision event, the maximum performance as a function of the number of CPU cores per GPU,   equal time comparisons of the VRC and TL estimators, and an analysis of ray casting in single versus double precision. 

\subsection{Test 1 -- 16x16 PWR fuel assembly} \label{pwrsection}
 
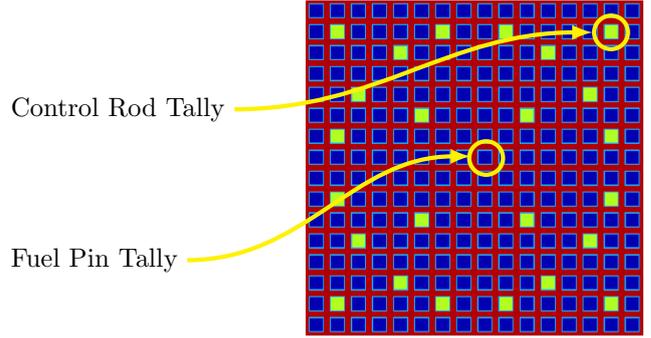
\begin{figure}[tbp] 
\begin{tikzpicture}
\node [anchor=west] (control) at (-1,3) {\normalsize Control Rod Tally};
\node [anchor=west] (fuel) at (-1,1) {\normalsize Fuel Pin Tally};
\begin{scope}[xshift=3.0cm]
    \node[anchor=south west,inner sep=0] (image) at (0,0) {\includegraphics[width=0.5\linewidth]{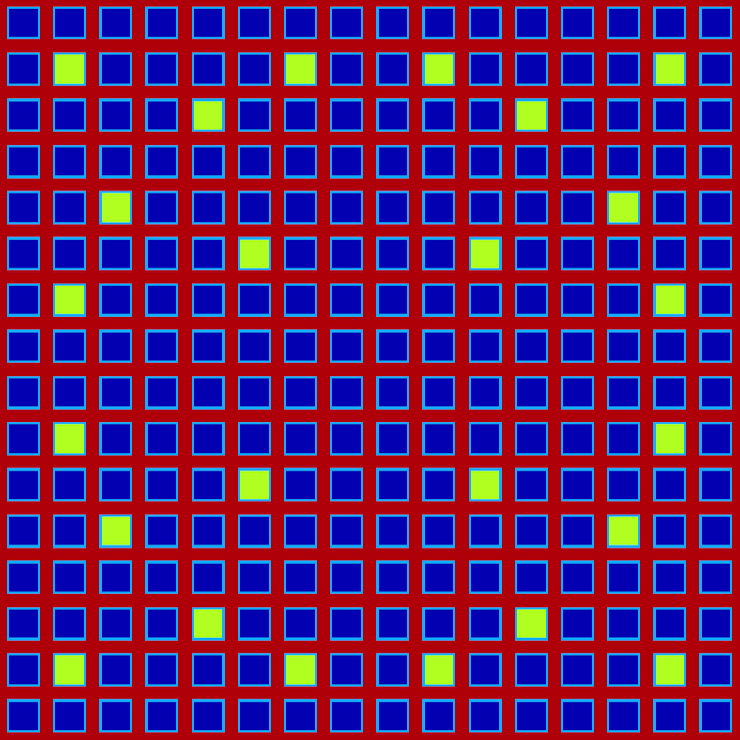}};
    \begin{scope}[x={(image.south east)},y={(image.north west)}]
        \draw[yellow,ultra thick,rounded corners] (0.905,0.905) circle (0.05);
        \draw[yellow,ultra thick,rounded corners] (0.535,0.53) circle (0.05);
        \draw [-latex, ultra thick, yellow] (control) to[out=0, in=180] (0.85,0.905);
        \draw [-latex, ultra thick, yellow] (fuel) to[out=0, in=180] (0.49,0.535);


    \end{scope}
\end{scope}
\end{tikzpicture}%
  \caption{PWR 16x16 fuel assembly with rectangular fuel pins (blue), control rods (yellow), and water moderator (red).   The performance of the \VRCE was tested for tallies within the central fuel pin and corner control rod indicated by the yellow circles.}
  \label{fig:assembly_diagram}
\end{figure}

A 16x16 PWR fuel assembly, converted to rectangular geometry by \cite{burke2015a}, was used to evaluate the \VRCE for the simulation of light water reactors (\autoref{fig:assembly_diagram}).   
As MonteRay does not yet have reflective boundaries, the assembly was placed within a solution of uranyl nitrate and this was reflected with graphite.  
The control rod pins consisted of B4C with 20 atomic percent Boron-10.   
The fuel was UO2 with 5 atomic percent enriched U-235.  
The geometry was specified with a 0.6 mm x 0.6 mm x 1000 mm mesh spacing within the assembly.   
The entire mesh, including the uranyl nitrate solution and graphite, was 340 x 340 x 5 cells.    
Total neutron fluence tallies were performed on the same mesh.  

\subsubsection{Performance vs. number of CPU cores}

\begin{figure*}[tbp]\centering 
\includegraphics[width=0.9\linewidth,trim=0.05cm 1.1cm 0.35cm 1.2cm,clip]{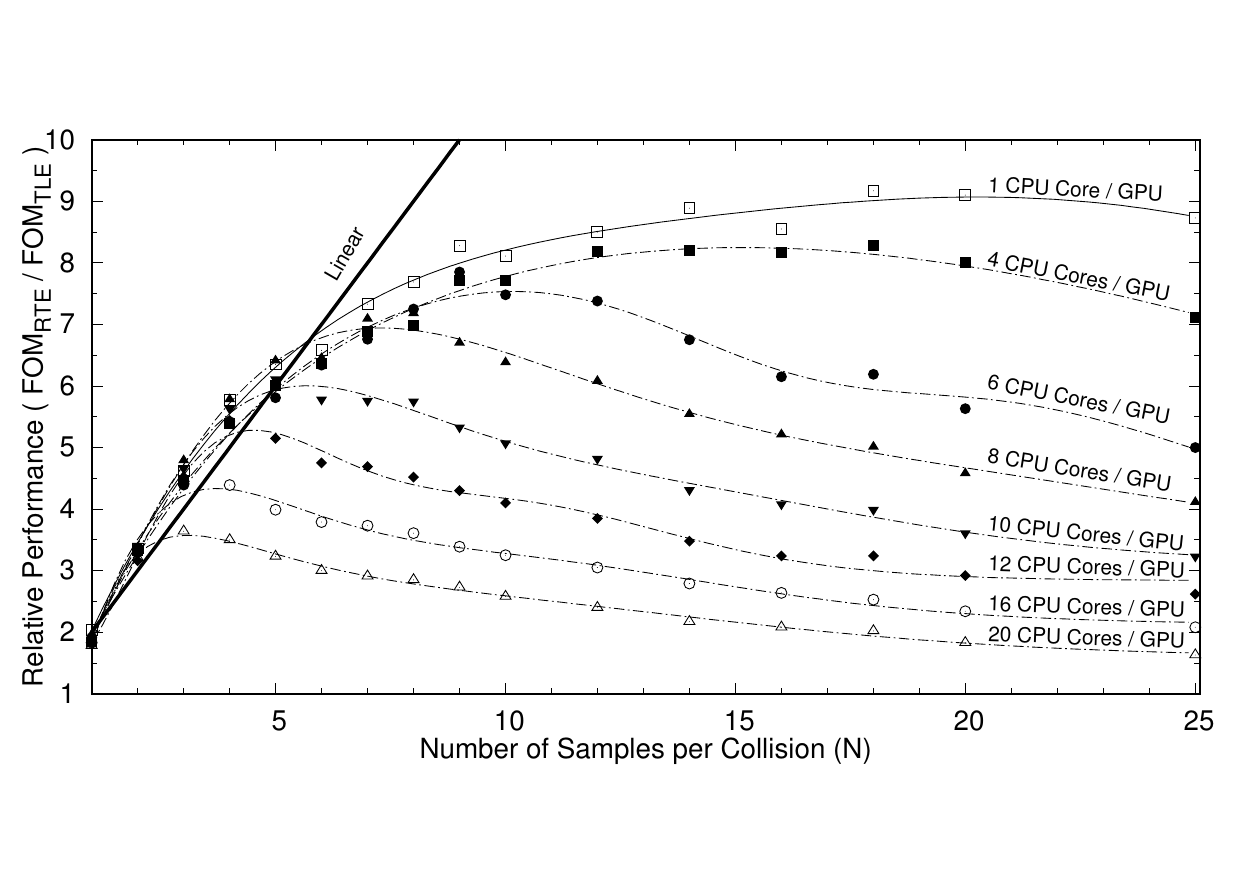}
\caption{Performance of the \VRCE for the corner control rod tally of the PWR 16x16 fuel assembly as a function of the number of ray samples per collision. }
\label{fig:pwr_fom_ratio_vs_nSample}
\end{figure*}

Two regions within the fuel assembly were evaluated to determine the effectiveness of the \VRCE: a central fuel pin and a corner control rod (\autoref{fig:assembly_diagram}).   
The average variances within these two regions was used to evaluate the performance of the \VRCEPERIOD.   
The performance was evaluated using MCATK's static \keff\ eigenvalue solution mode with 40,000 particles per cycle, 20 inactive cycles, and 40 active cycles.    
The performance was assessed as a function of both the number of rays sampled per collision and the number of CPU cores per GPU (\autoref{fig:pwr_fom_ratio_vs_nSample}).  
For a single CPU core per GPU, the \VRCE obtained maximum performance in the control rod using 18 ray samples per collision.   
For eight CPU cores per GPU, the maximum performance in the control rod was obtained using 7 ray samples per collision.   
The maximum performance was obtained when the GPU computation time matched the CPU computation time.  
The performance increase was approximately linear for small numbers of ray samples per collision, this indicates that in this range each ray sample was as effective as an independent random walk particle.  

\begin{figure*}[tbp]\centering 
\includegraphics[width=0.9\linewidth,trim=0.05cm 1.1cm 0.25cm 0.8cm,clip]{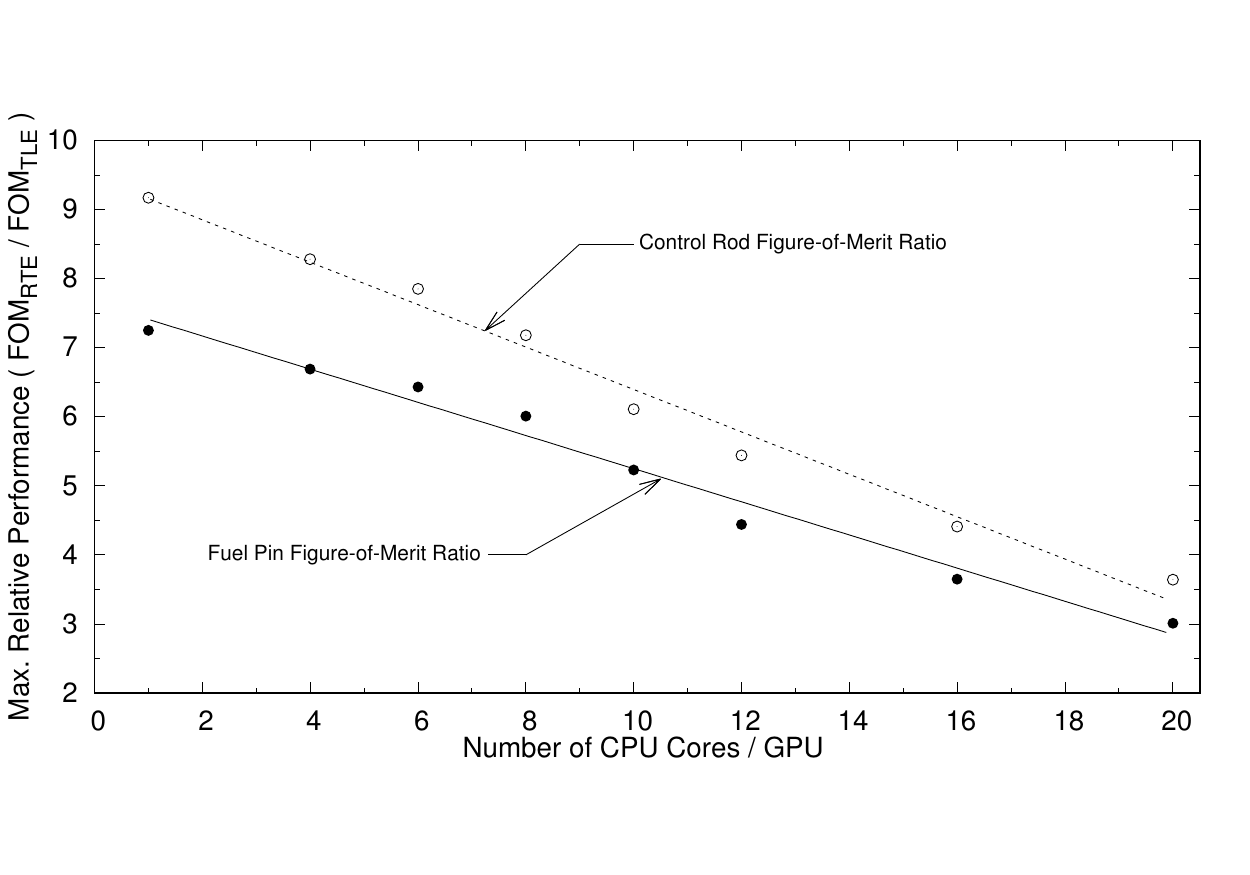}
\caption{Maximum performance of the \VRCE for PWR 16x16 fuel assembly as a function of the number of CPU cores matched to a GPU.}
\label{fig:pwr_fom_ratio_vs_nCores}
\end{figure*}

The maximum performance of the \VRCE is approximately a linear function of the number of CPU cores per GPU (\autoref{fig:pwr_fom_ratio_vs_nCores}).
The \VRCE obtained performance increases in the corner control rod of 9.2 and 7.2 for one and eight CPU cores per GPU respectively.  
In the fuel pin the performance increases were 7.3 and 6.0\@.    

\subsubsection{Equal time comparison}  
 
\begin{figure*}[ptb]
\centering
%
%
\subfigure[\TLE fluence  $( \text{n} \cdot  \left\{ \text{fission n} \right\}^{-1} \cdot \text{cm}^{-2} \cdot 10^{-4} )$) ][\raggedright \TLE fluence  $( \text{n} \cdot  \left\{ \text{fission n} \right\}^{-1} \cdot \text{cm}^{-2} \cdot 10^{-4} )$ ]{ 
  \includegraphics[width=0.36\linewidth]{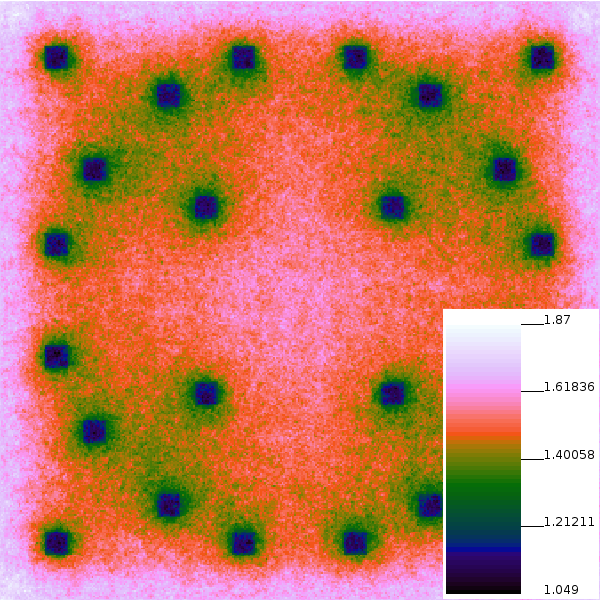}
  \label{fig:assembly_TL_fluence}  
}
%
\subfigure[\VRCE fluence  $( \text{n} \cdot  \left\{ \text{fission n} \right\}^{-1} \cdot \text{cm}^{-2} \cdot 10^{-4} )$) ][\raggedright \VRCE fluence  $( \text{n} \cdot  \left\{ \text{fission n} \right\}^{-1} \cdot \text{cm}^{-2} \cdot 10^{-4} )$ ]{ 
  \includegraphics[width=0.36\linewidth]{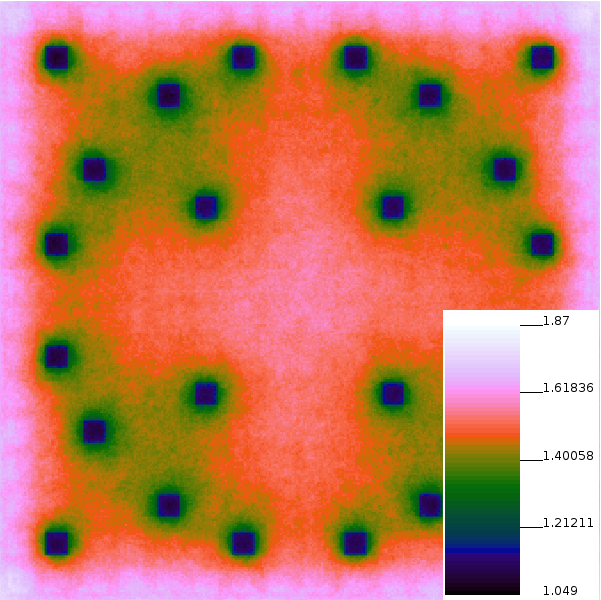}
  \label{fig:assembly_RT_fluence}  
}%

\subfigure[\TLE rel.\ uncertainty, $\sigma_{\text{TLE}}$ (\%)]{ 
  \includegraphics[width=0.36\linewidth]{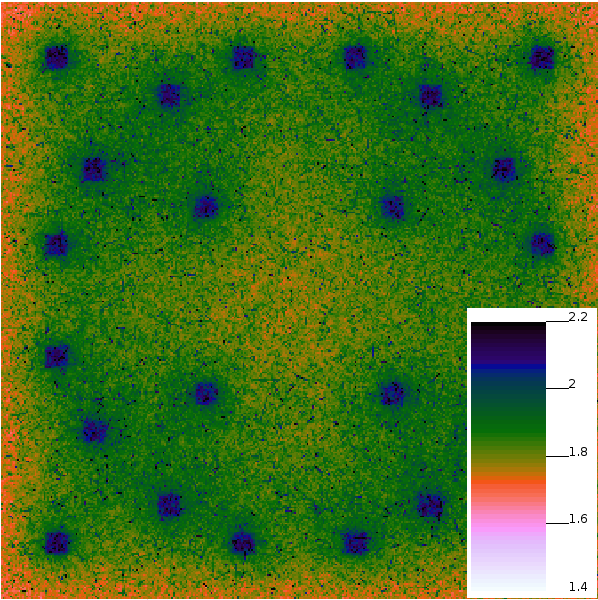}
  \label{fig:pwr_track_length_relError_plot}
}
%
\subfigure[\VRCE rel.\ uncertainty, $\sigma_{\text{VRCE}}$ (\%)]{ 
  \includegraphics[width=0.36\linewidth]{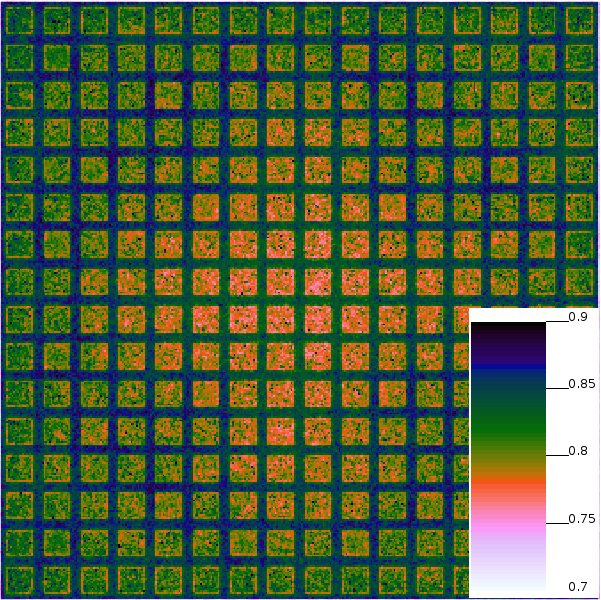}
  \label{fig:pwr_ray_trace_relError_plot}
}

\subfigure[\VRCE performance, $\eta$]{ 
  \includegraphics[width=0.36\linewidth]{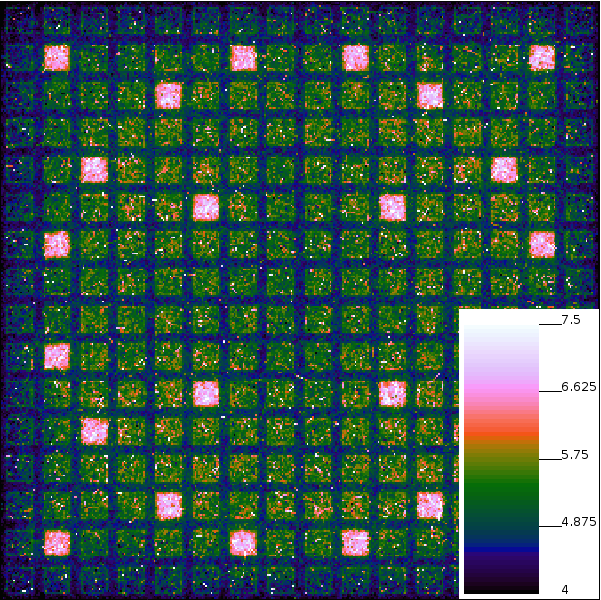}
  \label{fig:pwr_fom_ratio}
}
\caption{PWR 16x16 fuel assembly simulated with the \TLE and \VRCE for 600 seconds.  
The \TLE calculation used eight CPU cores with 40000 particles per cycle, 124 active cycles and 20 inactive cycles.  
The \VRCE used eight CPU cores paired with one NVIDIA Titan X GPU with 40000 particles per cycle, 93 active cycles, 20 inactive cycles, and 8 rays per collision. 
\TLE neutron fluence (a),  \VRCE  neutron fluence  (b),   \TLE relative uncertainty (c), \VRCE relative uncertainty  (d), \VRCE performance (e). }
\label{fig:pwr_plots}
\end{figure*}

In another test using the PWR assembly model, the \TLE and \VRCE were executed for equal wall clock times of 600 seconds. 
Thus, the ratio of time in Eq.~\eqref{eq:performance} was unity and the \VRCE performance was simply the ratio of the Monte Carlo variances.          
The \TLE calculation used 40,000 particles per cycle, 124 active cycles, and 20 inactive cycle on 8 CPU cores.
The \VRCE calculation used 40,000 particles per cycle, 93 active cycles, 20 inactive cycles, and 8 rays per collision with 8 CPU cores and a single GPU\@.
The \VRCE generated fluence results were generally less noisy than the \TLE results (Figs. \ref{fig:assembly_TL_fluence} and \ref{fig:assembly_RT_fluence}).   
The spatial distribution of the \VRCE uncertainty was considerably different and significantly lower than the \TLE uncertainty (Figs. \ref{fig:pwr_track_length_relError_plot} and \ref{fig:pwr_ray_trace_relError_plot}).   
While the \TLE estimator uncertainty was inversely proportional to the neutron fluence, the \VRCE uncertainty was not (note that the color scale of the uncertainty plots has been inverted from the color bar of the neutron fluence plots to show the inverse relationship).   
The \VRCE had the highest uncertainty in the water moderator and the \TLE estimator had the highest uncertainty in the control rods.     
The performance of the \VRCE has been calculated on a cell by cell basis (Fig. \ref{fig:pwr_fom_ratio}).   
The best performance increase was in the control rods and the lowest performance increase was in the water moderator.

The \VRCE calculation was also performed on the CPU with the same parameters as used on the GPU\@.   
However, instead of the 600 seconds, the CPU calculation took 5893 seconds on 8 CPU cores.  
Compared to the \TLE, this resulted in a \VRCE performance of 0.68\@.   
When using only the CPU for the \VRCE calculation, the additional computational expense usually outweighs the reduction in variance.         

\subsubsection{Single-precision vs. Double precision}

For the PWR fuel assembly, the maximum relative difference results calculated using double precision ray casting on a CPU and results calculated using single precision ray casting on a GPU ($\Delta_{\phi}$) was 0.0018\%. 
The maximum fractional difference ($\epsilon_{\phi}$) was 0.23\%.   
As the maximum fraction difference ($\epsilon_{\phi}$) was much less than 100\%, this indicated that the differences were not statistically significant.

\subsection{Test 2 -- Criticality Accident}

\begin{figure}[tbp]\centering 
\includegraphics[width=0.78\linewidth]{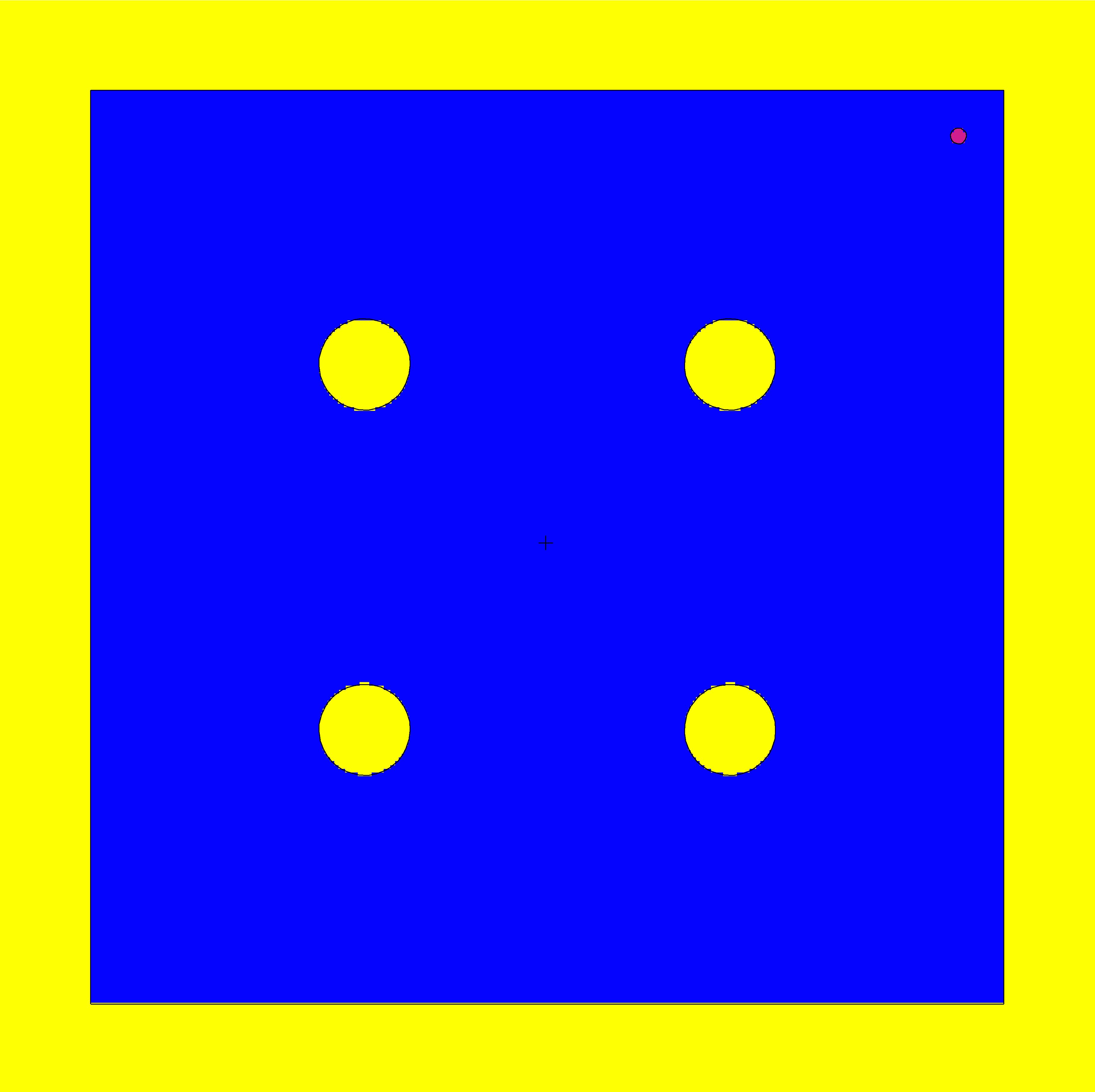}
\caption{Plan view of the geometry of the room in the criticality accident scenario.  A sphere of enriched uranium is in the top right corner(red).   The room has 1 m thick concrete walls and four 50-cm radius concert columns (yellow).   The room was filled with air (blue). }
\label{fig:criticality_accident_geometry}
\end{figure}

To evaluate the performance of the \VRCE in the context of radiation protection, a criticality accident has been simulated.    
The criticality accident was modeled with a sphere of U-235 placed in the corner of a concrete room.   
The interior room was 10 meters wide x 10 meters long x 2 meters high and filled with air (\autoref{fig:criticality_accident_geometry}).   
The concrete walls, floor, and ceiling of the room were 1 meter thick.   
Four 50-cm radius concrete columns were placed in the rooms.   
A mesh of 400 x 400 x 50 cells was used to model the room.   

\subsubsection{Performance vs. number of CPU cores}

To evaluate the performance of the \VRCE the total neutron fluence was calculated with the \VRCE and the \TLE using MCATK's \keff\ eigenvalue solution mode with 40,000 particles per cycle, 50 active cycles, and 5 inactive cycles.   
The average variance, in the mesh cells contained by a 50 cm radius sphere in the center of the room, was used to evaluate the performance of the \VRCEPERIOD.  

\begin{figure*}[tbp]\centering 
\includegraphics[width=0.9\linewidth,trim=0.05cm 1.1cm 0.2cm 1.2cm,clip]{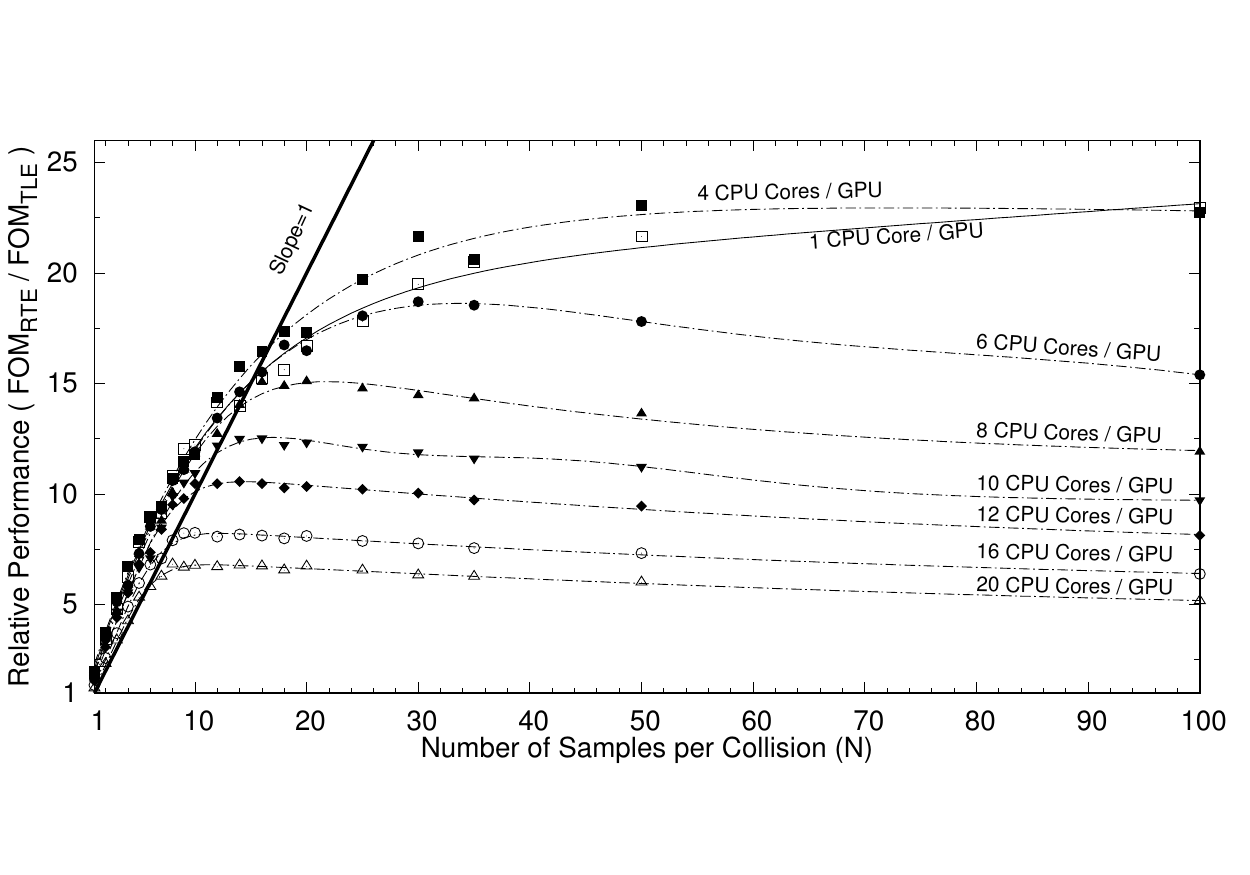}
\caption{Performance of the \VRCE as a function of the number of ray samples per collision for the criticality accident scenario. }
\label{fig:criticality_accident_fom_ratio_vs_nSample}
\end{figure*}

Again, the performance was assessed as a function of both the number of rays sampled per collision and the number of CPU cores per GPU (\autoref{fig:criticality_accident_fom_ratio_vs_nSample}).  
For a single CPU core per GPU, the \VRCE had the highest measured performance using 100 ray samples per collision.   
A larger number of rays per collision was not measured as larger number of CPU cores per GPU had maximum performance with less than 100 samples per collision.
For eight CPU cores per GPU, the maximum performance was obtained using 20 ray samples per collision.     

\begin{figure*}[tbp]\centering 
\includegraphics[width=0.9\linewidth,trim=0.0cm 1.0cm 0.5cm 0.8cm,clip]{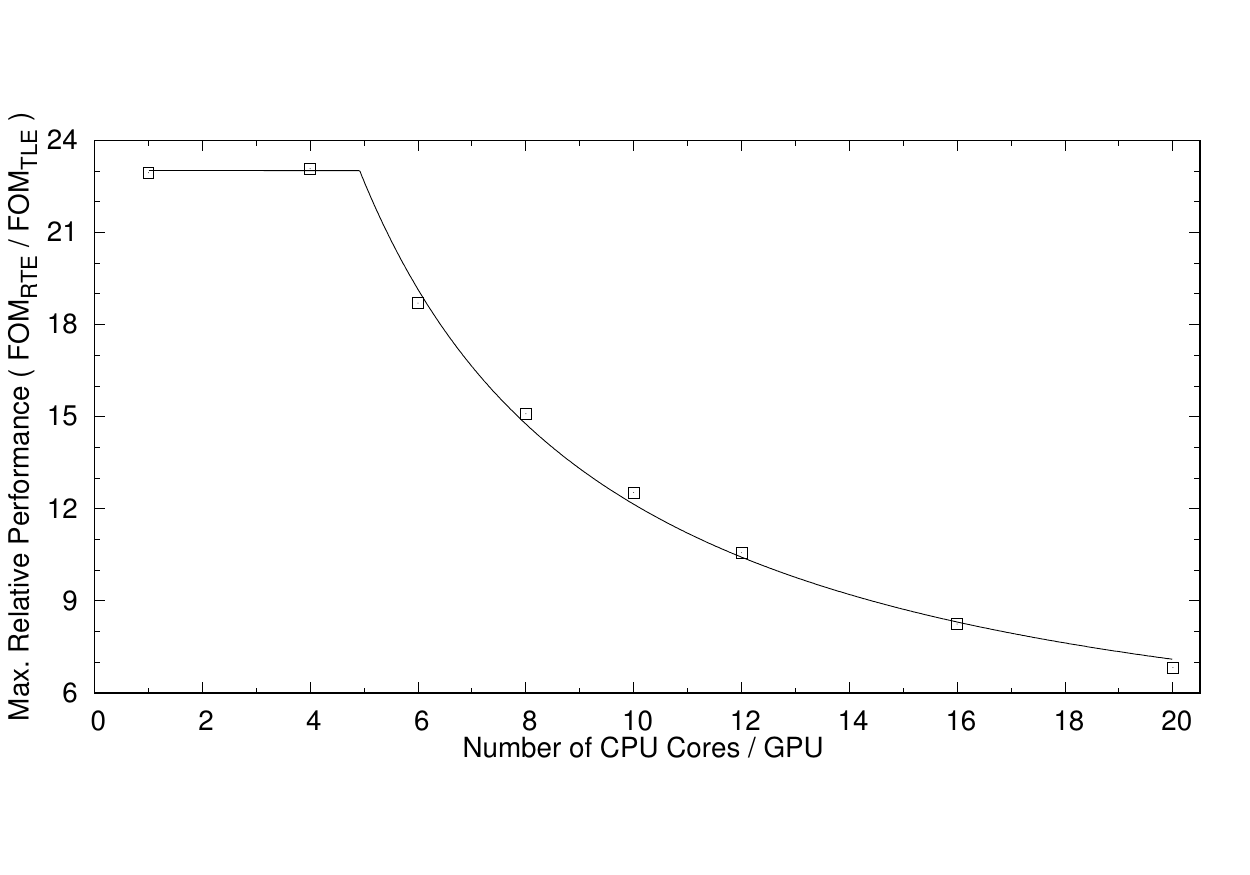}
\caption{Maximum performance of the \VRCE as a function of the number of CPU cores per GPU for the criticality accident scenario.}
\label{fig:accident_fom_ratio_vs_nCores}
\end{figure*}

The maximum performance of the \VRCE  is constant up to four CPU cores per GPU (\autoref{fig:accident_fom_ratio_vs_nCores}).  
With more than four CPU cores per GPU the performance is inversely proportional to the number of CPU cores.   
The performance increase, as measured in the center of the room, was 23.0 and 15.1 for one and eight CPU cores per GPU respectively.  

The \VRCE performance on the CPU (as opposed to calculating the \VRCE on the GPU) was also assessed as a function of the number of rays sampled per collision.
Compared to the \TLE, the \VRCE on the CPU had a maximum performance increase of 1.24 times with 12 ray samples per collision.   This indicates that this method may be useful for some problem types on conventional CPU hardware without GPU acceleration.

\subsubsection{Equal time comparison}   

\begin{figure*}[ptb]
\centering
\subfigure[\TLE fluence  $( \text{n} \cdot  \left\{ \text{fission n} \right\}^{-1} \cdot \text{cm}^{-2} \cdot 10^{-4} )$) ][\raggedright \TLE fluence  $( \text{n} \cdot  \left\{ \text{fission n} \right\}^{-1} \cdot \text{cm}^{-2} \cdot 10^{-4} )$ ]{ 
  \includegraphics[width=0.36\linewidth]{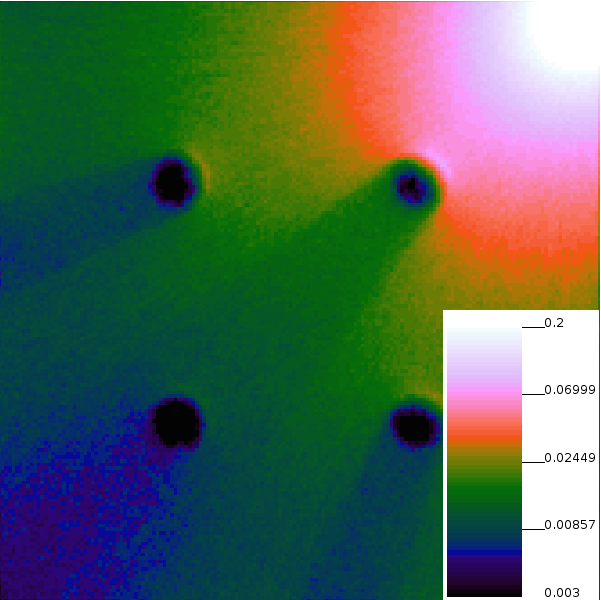}
  \label{fig:accident_tl_fluence}  
}
%
\subfigure[\VRCE fluence  $( \text{n} \cdot  \left\{ \text{fission n} \right\}^{-1} \cdot \text{cm}^{-2} \cdot 10^{-4} )$) ][\raggedright \VRCE fluence  $( \text{n} \cdot  \left\{ \text{fission n} \right\}^{-1} \cdot \text{cm}^{-2} \cdot 10^{-4} )$ ]{
  \includegraphics[width=0.36\linewidth]{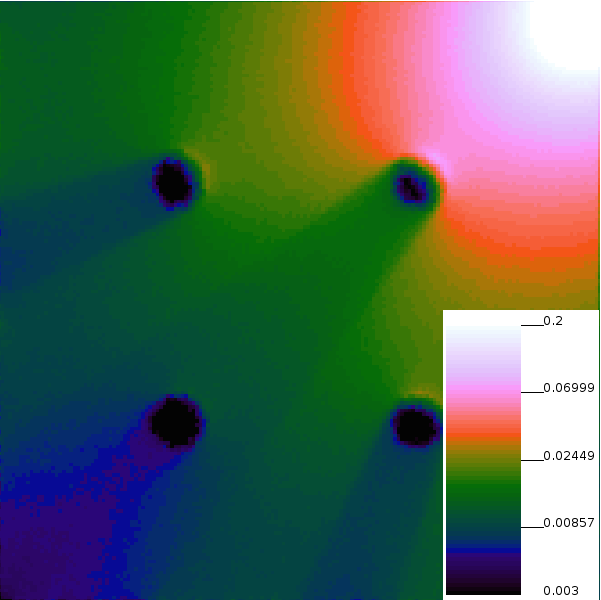}
  \label{fig:accident_rt_fluence}  
}
\subfigure[\TLE rel.\ uncertainty, $\sigma_{\text{TLE}}$ (\%) ]{ 
  \includegraphics[width=0.36\linewidth]{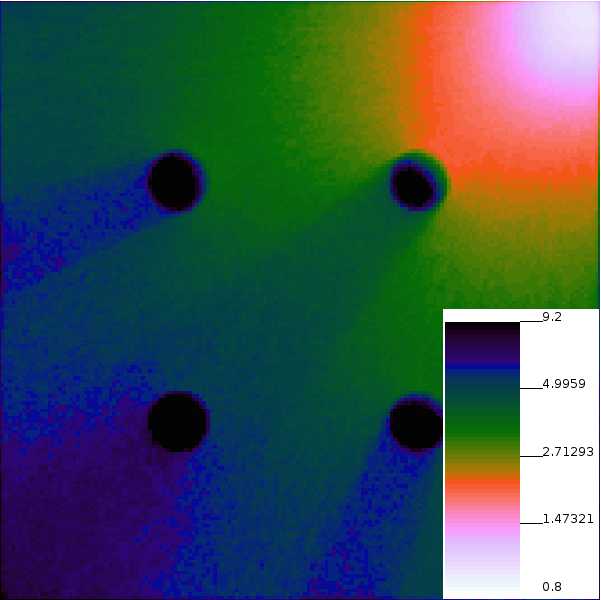}
  \label{fig:accident_tl_error}
}
%
\subfigure[\VRCE rel.\ uncertainty, $\sigma_{\text{VRCE}}$ (\%)]{ 
  \includegraphics[width=0.36\linewidth]{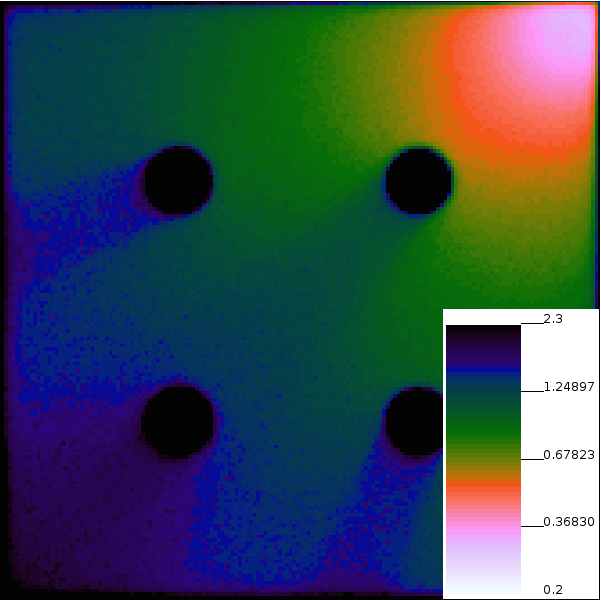}
  \label{fig:accident_rt_error}
}

\subfigure[\VRCE performance, $\eta$]{ 
  \includegraphics[width=0.36\linewidth]{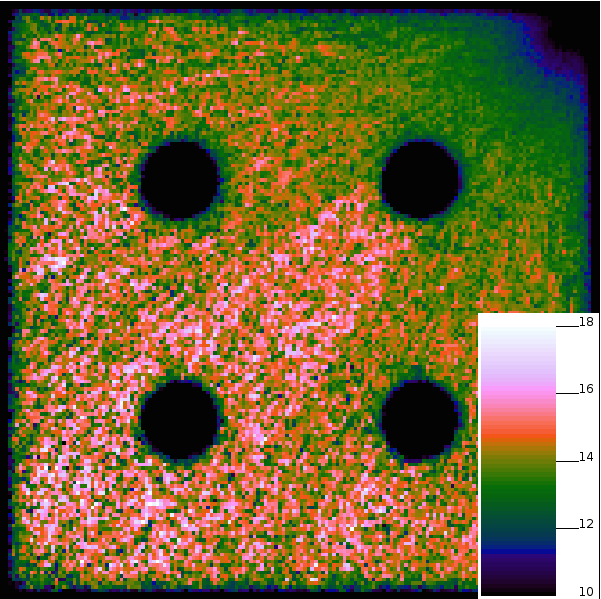}
  \label{fig:accident_fom_ratio}
}
%
\caption{Neutron fluence in a concrete room due to a criticality accident calculated with the \VRCE and \TLE using equal computation times.   
Both calculations had a wall clock time of 645 seconds.    
The \TLE calculation used eight CPU cores with 40000 particles per cycle, 151 active cycles and 5 inactive cycles.  
The \VRCE used eight CPU cores paired with one NVIDIA Titan X GPU with 40000 particles per cycle, 63 active cycles, 20 inactive cycles, and 20 rays per collision. 
\TLE neutron fluence (a),  \VRCE  neutron fluence  (b),   \TLE relative uncertainty (c), \VRCE relative uncertainty  (d), \VRCE performance (e). } 
\label{fig:accident_plots}
\end{figure*}

The criticality accident scenario was simulated with both the TL and VRC estimators for equal time (645 seconds) using 8 CPU cores.   
The \TLE calculation used 40,000 particles per cycle, 151 active cycles, and 5 inactive cycles.   
The \VRCE calculation used 40,000 particles per cycle, 63 active cycles, 5 inactive cycles, and 20 rays per collision.   
The \VRCE produced a much smoother neutron fluence than the \TLE (Figs. \ref{fig:accident_tl_fluence} and \ref{fig:accident_rt_fluence}).  
Rays from individual particles are clearly seen in the \TLE fluence, but have been smoothed out in the \VRCE fluence.  
The neutron fluence shadows, created by the concrete columns, have sharp edges when using the \VRCE that are not as clearly defined with the \TLEPERIOD.
The Monte Carlo uncertainty of the \VRCE and the \TLE follow the same general trends (Figs. \ref{fig:accident_tl_error} and \ref{fig:accident_rt_error}).
However, the uncertainty of the \VRCE is generally 4 times lower than the uncertainty of the \TLE (the color scale of the \TLE is 4 times the color scale of the \VRCE). 
As variance is the square of the uncertainty, this results in a performance increase of 16.     
A plot ratio of the \VRCE performance indicates that performance was generally 12 - 18 for locations in air (\autoref{fig:accident_fom_ratio}).   

\subsubsection{Single-precision vs. Double precision}

The \VRCE fluence values calculated with single-precision ray casting on the GPU was again compared to \VRCE fluence values calculated with double-precision on the CPU\@.
As the cells in the concrete columns have very poor statistics, as much as 100\% uncertainty, it is not surprising to find some differences in these fluence values.   
There were 3 cells out of 2 million cells that had relative differences that were larger than 1\% but none that had differences larger than the uncertainty. 
The uncertainty in these 3 cells was extremely poor, greater than 71\%\@.   
6 other cells had only a single particle track contribute to their fluence estimate and also had relative differences greater than 1\% but the uncertainty is not defined for a single sample.   
These differences seem to indicate slight ray casting differences, which is to be expected even for double precision implementations on different hardware. 
More meaningful results can be obtained by only examining the differences in cells with statistically significant fluence tally estimates, cells with uncertainties less then 5\%\@.    
The maximum relative difference ($\Delta_\phi$) was 0.0011\% for these cells.  
The maximum fraction difference ($\epsilon_\phi$) was also very low, less than 0.1\%\@.

\subsection{Test 3 -- Reflected Godiva Criticality Benchmark}

Simulation of the Reflected Godiva criticality benchmark~\citep{GodivaR} was performed to demonstrate the performance of the \VRCE in optically thick systems.  
The Reflected Godiva benchmark was modeled as a 6.5 cm radius, highly enriched, uranium metal core surrounded by a sphere of water with a 33.3 cm radius.   
The geometry was modeled with a Cartesian mesh using 100 x 100 x 100 cells.  

\subsubsection{Performance vs. number of CPU cores}

\begin{figure*}[tbp]\centering 
\includegraphics[width=0.9\linewidth,trim=0.0cm 1.1cm 0.4cm 1.2cm,clip]{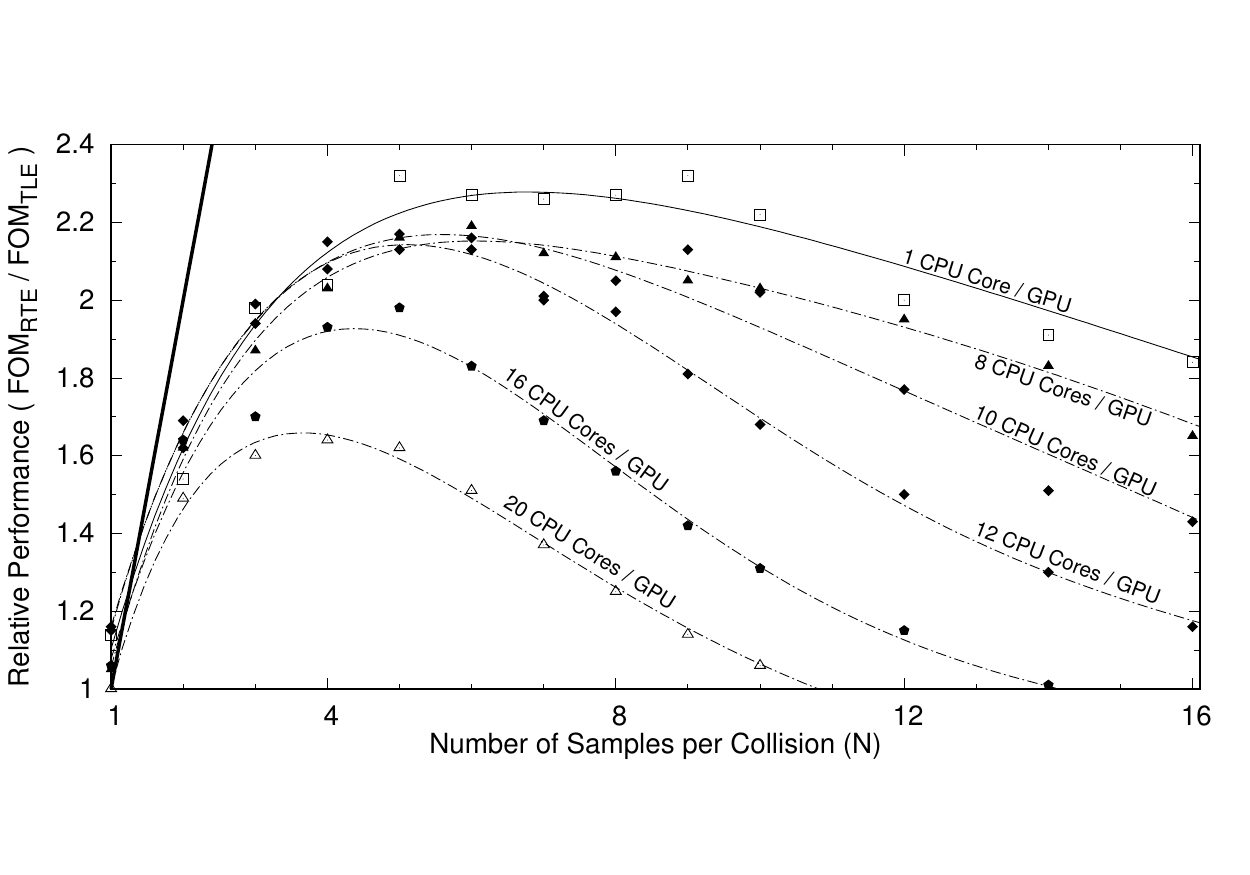}
\caption{Performance of the \VRCE as a function of the number of ray samples per collision for Reflected Godiva.  Performance was measured using the average variance in a 1 cm radius spherical region located at the center of the metal core.}
\label{fig:godivaR_variance_ratio}
\end{figure*}

The performance of the \VRCE was assessed using 40,000 particles per cycle, 25 active cycles, and 10 inactive cycles.  
The average variance in a 1 cm radius sphere at the center of the metal core was used to calculate the performance (\autoref{fig:godivaR_variance_ratio}).    
Using a single CPU core per GPU, the \VRCE had the highest measured performance using between 5 and 9 rays per collision (5 and 9 rays per collision provided the same FOM).   
Using eight CPU cores per GPU, the maximum performance was obtained using 6 rays per collision.    

\begin{figure*}[tbp]\centering 
\includegraphics[width=0.9\linewidth,trim=0.0cm 1.1cm 0.4cm 1.0cm,clip]{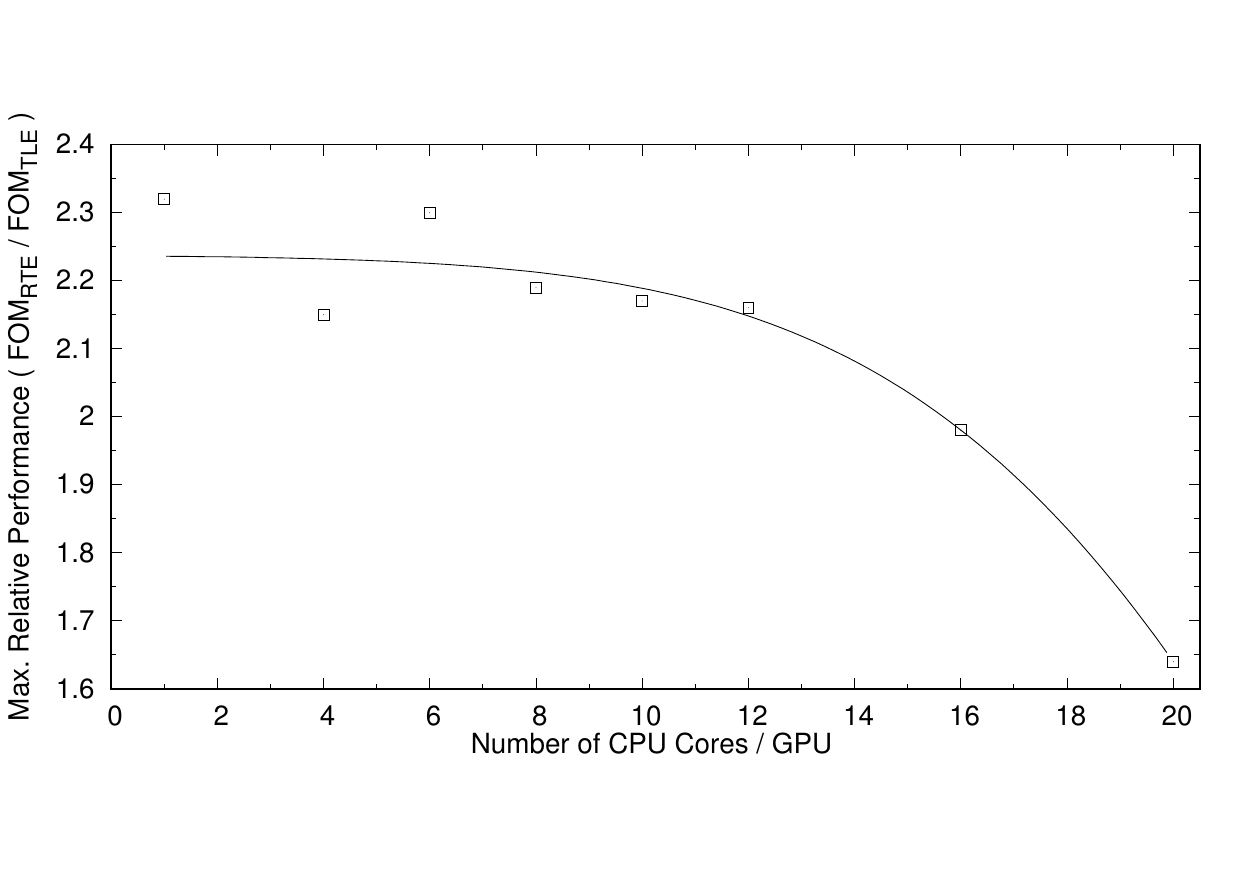}
\caption{Performance of the \VRCE as a function of the number of CPU cores per GPU for Reflected Godiva.}
\label{fig:godivaR_fom_ratio_vs_nCores}
\end{figure*}

The maximum performance of the \VRCE as a function of the number of CPU cores was relatively constant up to 12 cores (\autoref{fig:godivaR_fom_ratio_vs_nCores} ).   
The maximum performance was 2.3 and 2.2 for 1 and 8 cores respectively. 
For this test problem, the performance of the \VRCE is limited by method and not by the ray casting rate of the GPU.  

\subsubsection{Equal time comparison}   

\begin{figure}[bpt]
\centering
  \includegraphics[width=0.78\linewidth]{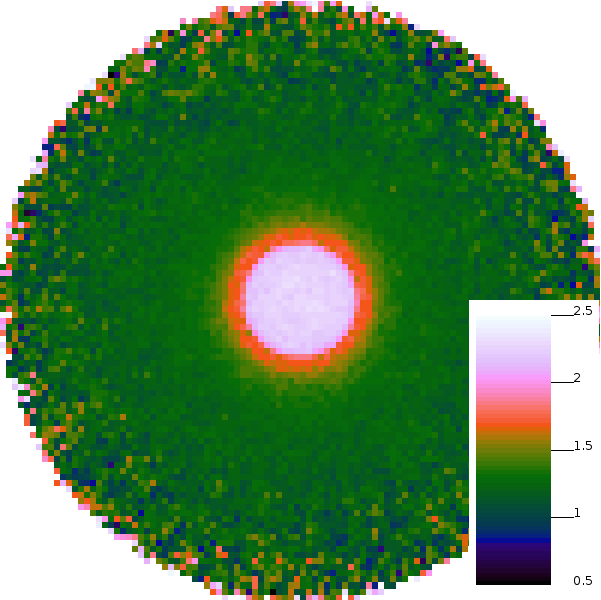}
  \caption{\VRCE performance for simulations of equal time (600 seconds) for the Reflected Godiva benchmark.  Both the \VRCE and \TLE calculations used 8 CPU cores. }
  \label{fig:godivaR_fom_ratio}
\end{figure}

The reflected Godiva criticality benchmark was simulated using both methods for 600 seconds with 8 CPU cores.   
The calculation with the \TLE used 40,000 particles per cycle, 241 active cycles, and 10 inactive cycles.    
The calculation with the \VRCE used 40,000 particles per cycle, 168 active cycles, 5 inactive cycles, and 6 rays per collision.     
The fluence and Monte Carlo uncertainty distributions were similar for these two calculations.   
Only the magnitude of the uncertainty was different.   Thus the plots of the fluence and the uncertainty have not been included.  
The performance of the \VRCE within the water reflector is approximately one, while the performance within the Uranium metal is between 2.0 and 2.5 (\autoref{fig:godivaR_fom_ratio}).

\subsubsection{Single-precision vs. Double precision}

The maximum relative difference in single-precision results on the GPU and double precision results on the CPU for the 1 million cells was 0.14\%.   
However, there were many cells with large uncertainties in the outer edge of the water reflector.
The maximum relative difference ($\Delta_\phi$) in the cells with statistically significant fluence estimates, less than 5\% uncertainty, was 0.0015\%.  
The maximum fraction difference ($\epsilon_\phi$) was less than 0.07\%.

\section{Discussion}

\subsection{Review of Other Research}

The \VRCE method does not attempt to provide the best possible performance of Monte Carlo particle transport on GPU hardware.  
It is expected that the best possible performance is attainable by porting the Monte Carlo random walk to the GPU\@.    
In order to understand how the \VRCE method compares to porting the entire Monte Carlo algorithm to the GPU, a review of other research has been performed.
Comparisons are difficult as other researchers may use different hardware and have significantly different capabilities in their Monte Carlo GPU implementations.   

One difficulty that arises from such comparisons is that researchers often quote their GPU performance relative to single core CPU performance.
While such numbers are often impressive they are extremely misleading.   
The day of the single core CPU compute node has long since passed.   
Moore's Law is now expressed by the increase in the number of cores not in an increase in the CPU clock speed.   
Today dual and quad core laptops, tablets, and cell phones are common.   
But it is relatively easy to convert GPU performance relative to single core CPUs to multiple core CPUs\@.   
The performance of the Monte Carlo codes, like MCATK and \MCNP, scales linearly for the relatively small number of cores to be found on a single compute node.   
This performance can be expressed as 

\begin{equation}
P_{CPU}(n) = 1 + (n-1)  \epsilon
\end{equation}
where $n$ is the number of cores and $\epsilon$ is the strong scaling efficiency.   
For the three test problems used in this study MCATK has a strong scaling efficiency of $\sim$78\% for up to 20 cores.    
The performance of multiple CPU cores can then be used to scale the measurements reported by other researchers:
\begin{equation}
P_{GPU}(n) = \frac{P_{GPU}(1) }  {P_{CPU}(n) }=  \frac{P_{GPU}(1)} { 1 + (n-1)  \epsilon}
\label{eq:gpu_ncpus}
\end{equation}
where $P_{GPU}(n)$ is the performance of a single GPU compared to the performance of $n$ CPU cores. 

Another problem that is common with other researchers implementations is that the physics is oversimplified.   
For example several researchers used one, or few, group neutron cross-sections.   
This does not adequately test the cross-section lookup on the GPU and this lookup can account for a significant percentage of the computational costs.   

\cite{wolfeArcher} used the ARCHER code to demonstrate neutron transport with 1 group cross-sections.  
They tested ARCHER on an Intel Xeon X5650 CPU at 2.66GHz and on four different GPUs.
The GPUs tested include an NVIDIA M2090, NVIDIA GTX Titan, NVIDIA K20, and an NVIDIA K40\@.   
They obtained a 6.84 times speed up on the K40 GPU over a six core CPU for a 1-D slab problem. 
Their performance comparison with several GPU models is helpful in scaling other researchers results reported using a single GPU model.
They reported that the NVIDIA K40 was 1.43 times faster than the NVIDIA K20.  
The K40 to K20 performance ratio of 1.43 will be used to scale the results from other researchers that used less powerful GPUs\@ than the K40.

\cite{nelson2009} developed a simplified Monte Carlo neutron transport code for GPU hardware.  
Nelson's physics models were basic.   
The cross-sections contained a few hundred energy points and  the geometry was 3-D solid body geometry using only spheres and parallelepipeds. 
Nelson's work provides a good expectation of what is possible if a production code is ported to GPU hardware, obtaining a 23.9 times speedup compared to a single CPU\@.
Testing was performed with a NVIDIA GeForce GTX 275 GPU and an Intel Core i7-920 CPU at 2.66 GHz.
If Nelson's tests were rerun on modern GPU hardware the performance improvement may be higher.
Scaling Nelson's results by the factor of 1.43 results in a 34.1 times speedup compared to a single CPU\@.
This scale factor is probably low for the GTX 275, but it is desired to use a single scaling factor for all non-K40 results.  

\cite{Xu2013} used RMC to demonstrate realistic neutron transport on a GPU\@.   
They used a cylindrical (R-Z) mesh and 30 group nuclear data.   
Testing of RMC was performed on a NVIDIA Geforce GTX680 in comparison with a four core Intel Q9400 CPU at 2.66 GHz.
They found a performance increase of 5.6 times.   
The GTX680 performance is nearly equivalent to the Nvidia Tesla K20~\citep{gpufaceoff}, therefore using the K20 to K40 scale factor is appropriate. 
Scaling the RMC results by the factor of 1.43 results in a 8.0 times speedup compared to a quad core CPU\@.

\cite{hamilton2016} used the Profugus mini-app, derived from Oak Ridge National Laboratory's Shift Monte Carlo code~\citep{shift}, to evaluate neutron transport on GPUs.  
They evaluated complex models including a 5x5x5 checkerboard problem of alternating moderator and 4\% enriched fuel using 56 group neutron cross-sections.
This problem is similar to the PWR assembly evaluated with MonteRay.  
The evaluation was performed using an eight core Intel Xeon E5-2630 CPU at 2.40 GHz and a NVIDIA Tesla K40 GPU\@.
They were able to obtain an increase in performance of 3.6 as compared to Profugus on an eight core CPU\@.    
Given its geometric representation, physics fidelity, and direct comparison to CPU performance, Profugus is probably the best GPU implementation with which to gauge the performance of MonteRay. 

\cite{bleile2016b} used Livermore National Laboratory's ALPSMC code, which models particle transport in 1-D binary stochastic media, to compare event-based and history based  implementations for GPU hardware.
They obtained a speedup of 54.62 using 2 Nvidia Tesla K80 GPUs compared to a single core of an Intel Xeon Haswell CPU at 3.2 GHz.     

\cite{bergmann2017} has probably developed the most full featured Monte Carlo code specifically designed for GPU hardware, WARP\@.  
Like MonteRay, WARP uses continuous energy ACE cross-section data.  
The ray tracing in WARP is performed by Nvidia's Optix~\citep{optix2} library.  
 They tested WARP against \MCNP~\citep{mcnp6} and Serpent~\citep{serpent} using complex reactor like geometries.  
 They analyzed 6 problems, but a hexagonal pin cell lattice is most similar to the PWR assembly used to test MonteRay.
 It was also the most complex problem they analyzed. 
For this problem they found a 3.22 times speedup with WARP on an Nvidia Titan Black as compared to \MCNP using 24 cores of 2 Intel Xeon E5-2670 v3 CPUs running at 2.3 GHz. 
And they obtained a 1.27 times speedup of WARP as compared to Serpent.
For the hexagonal pin cell lattice, they obtained near identical results running on an Nvidia Tesla K80. 
The best comparison for gauging GPU to CPU performance is the comparison to Serpent.     
 The speed of Serpent  (44.32 minutes) vs \MCNP (112.25 minutes) shows that Serpent is a fairer benchmark of GPU/CPU performance.
 There is no direct comparison of WARP running on CPU hardware.  
 Bergmann's early research made some comparisons between CPU and GPU performance~\citep{bergmann2014}.  
 He performed some tests of transport in 2D with mono-energetic scattering.   
With this scattering test he obtained a speedup of 14 on a Telsa K20 GPU over serial timings on an AMD Opteron 6128 at 2.0 GHz\@.   
Scaling Bergmann's results using the relative performance of the K40 to the K20 ($1.43\times$) results in a projected speed of 20.0\@.
 
The performance results from ALPSMC, ARCHER, Bergmann, Nelson, Profugus, RMC, and WARP are listed in \autoref{tab:gpu_cpu_data} along with the results of MonteRay\@.    
The results are normalized using a linear scaling to a 2.60 GHz CPU clock speed.   
The results are also scaled to a single CPU core and eight CPU cores using Eq.~\eqref{eq:gpu_ncpus} and a 78\% strong scaling efficency\@. 
Bergmann's direct comparison to CPU results \citep{bergmann2014} have been referred to as ``Bergmann2014'', while WARP to Serpent comparisons  \citep{bergmann2017}  are referred to as ``WARP vs. Serpent''\@.

\begin{table*}[tbp]
\begin{center}
\begin{tabular}{p{4.1cm}x{2.0cm}x{0.8cm}x{0.8cm}y{1.2cm}y{1.3cm}y{1.3cm}y{1.3cm}}
\toprule
\begin{minipage}[b]{3.1cm}\setlength{\baselineskip}{0.8\baselineskip}  Author/Code  \end{minipage} &
\begin{minipage}[b]{1.9cm}\setlength{\baselineskip}{0.8\baselineskip}  \centering GPU  \end{minipage} &
\begin{minipage}[b]{0.8cm}\setlength{\baselineskip}{0.8\baselineskip}  \centering \# of CPU Cores Used  \end{minipage} &
\begin{minipage}[b]{0.8cm}\setlength{\baselineskip}{0.8\baselineskip}  \centering CPU Clock (GHz)  \end{minipage} &
\begin{minipage}[b]{1.2cm}\setlength{\baselineskip}{0.8\baselineskip}  \flushright GPU Speedup  \end{minipage} &
\begin{minipage}[b]{1.3cm}\setlength{\baselineskip}{0.8\baselineskip}  \flushright GPU Speedup Scaled to  \mbox{2.60 GHz} \end{minipage} &
\begin{minipage}[b]{1.3cm}\setlength{\baselineskip}{0.8\baselineskip}  \flushright GPU Speedup vs. \mbox{Single} CPU Core @ \mbox{2.60 GHz} \end{minipage} &
\begin{minipage}[b]{1.3cm}\setlength{\baselineskip}{0.8\baselineskip}  \flushright GPU Speedup vs. Eight  CPU Cores @ \mbox{2.60 GHz} \end{minipage} \\
\midrule
RMC                      &  GTX680       & 4                      &  2.66   &  5.6              &                   &                         &                     \\
                              &  K40 Scaled\textsuperscript{*}  &    &     &  8.0              &  7.8            &    26.0                &  4.0                     \\
ALPSMC               & Tesla K80      & 0.5                   &  3.20   &  54.62         & 44.4            &   24.9                 &  3.9                        \\
Profugus               & Tesla K40      & 8                      &  2.40   &  3.6             & 3.9              &   25.2                 &  3.9                        \\
Bergmann2014     & Telsa K20      & 1                      &  2.00   &  14              &                    &                           &                        \\
                              &  K40 Scaled\textsuperscript{*}  &    &     &   20.0         & 25.9            &    25.9                 &  4.0                     \\
WARP vs. Serpent  & Titan Black    & 24                    & 2.30    &  1.27         & 1.44            &   27.2                 &  4.2                       \\
ARCHER               & Tesla K40     & 6                      &  2.66   &  6.84           & 6.7              &  32.8                 &  5.1                        \\
Nelson                   & GTX 275       & 1                      &  2.66   &  23.9            &                   &                  &                                 \\
                              &  K40 Scaled\textsuperscript{*}  &    &     &  34.1            &  33.3          &    33.3                &  5.2                     \\

MonteRay \\
\hfill \begin{minipage}[t]{3.1cm}\setlength{\baselineskip}{0.7\baselineskip}  Reflected \mbox{Godiva}  \end{minipage}  & Titan X     & -   &  2.60 & - & -  & 2.3 & 2.2 \\ 
\hfill \begin{minipage}[t]{3.1cm}\setlength{\baselineskip}{0.7\baselineskip}  PWR Fuel Pin                    \end{minipage} & Titan X     & -   &  2.60 & - & - &  7.3 & 6.0  \\
\hfill \begin{minipage}[t]{3.1cm}\setlength{\baselineskip}{0.7\baselineskip}  PWR Control Rod              \end{minipage} & Titan X     & -   &  2.60 & - & - &  9.2 & 7.2  \\
\hfill \begin{minipage}[t]{3.1cm}\setlength{\baselineskip}{0.7\baselineskip}  Criticality \mbox{Accident} \end{minipage} & Titan X     & -   &  2.60 & - & - & 23.0 & 15.1 \\
\bottomrule 
\multicolumn{8}{l}{\textsuperscript{*}\footnotesize{Scaled to the K40 using a factor of 1.43 from \cite{wolfeArcher}.}}
\end{tabular} 
\end{center}
\caption{The performance increases of GPU Monte Carlo codes obtained by other researchers compared to the results obtained with the 3 MonteRay test problems. 
The projection to one and eight cores has been performed using Eq.~\eqref{eq:gpu_ncpus}\@ and a strong scaling efficiency of 78\%.  
}
\label{tab:gpu_cpu_data}
\end{table*}%

The results of ``WARP vs. Serpent'' and MonteRay were not scaled to the K40 as they had better performance on Titan Black and Titan X GPUs\@.   
The ALPSMC data obtain on a Tesla K80 was also not scaled to the K40\@. 
The purpose of this compilation of data was not to provide exact comparisons.   
This could only be possible if each code is benchmarked on exactly the same hardware.   
Instead the purpose of this compilation was to indicate, using data from multiple implementations, what the upper bound might be for the performance of modern GPUs relative to a modern multi-core CPU\@.   
The results scaled from a single core to eight cores might be significantly in error.   
But the RMC, ARCHER, Profugus, and ``WARP vs. Serpent'' results, which were obtained with four, six, eight, and 24 CPU cores respectively, seem to validate the scaling of the Nelson, ALPSMC, and Bergmann2014 data.

After normalizing the results of the other researchers it is quite surprising to see that all seven of the other researchers results are very similar. 
The average speedup was 4.4 as compared to 8 CPU cores running at 2.60 GHz\@.  
ARCHER and Nelson had the largest performance gain of these five results with a speedup of 5.1 and 5.2 respectively\@.  
These results seem to indicate a performance wall of $\sim$4-5 times speedup of Monte Carlo neutron transport on modern GPUs as compared to eight CPU cores.  
Or more precisely, there seems to be a performance wall when the neutron random walk is ported to GPUs\@.   
The \VRCE tally shows that there are other methods for increasing performance. 

\begin{figure*}[tb]
\centering
  \includegraphics[width=1.0\linewidth,trim=0.0cm 1.0cm 0.3cm 0.8cm,clip]{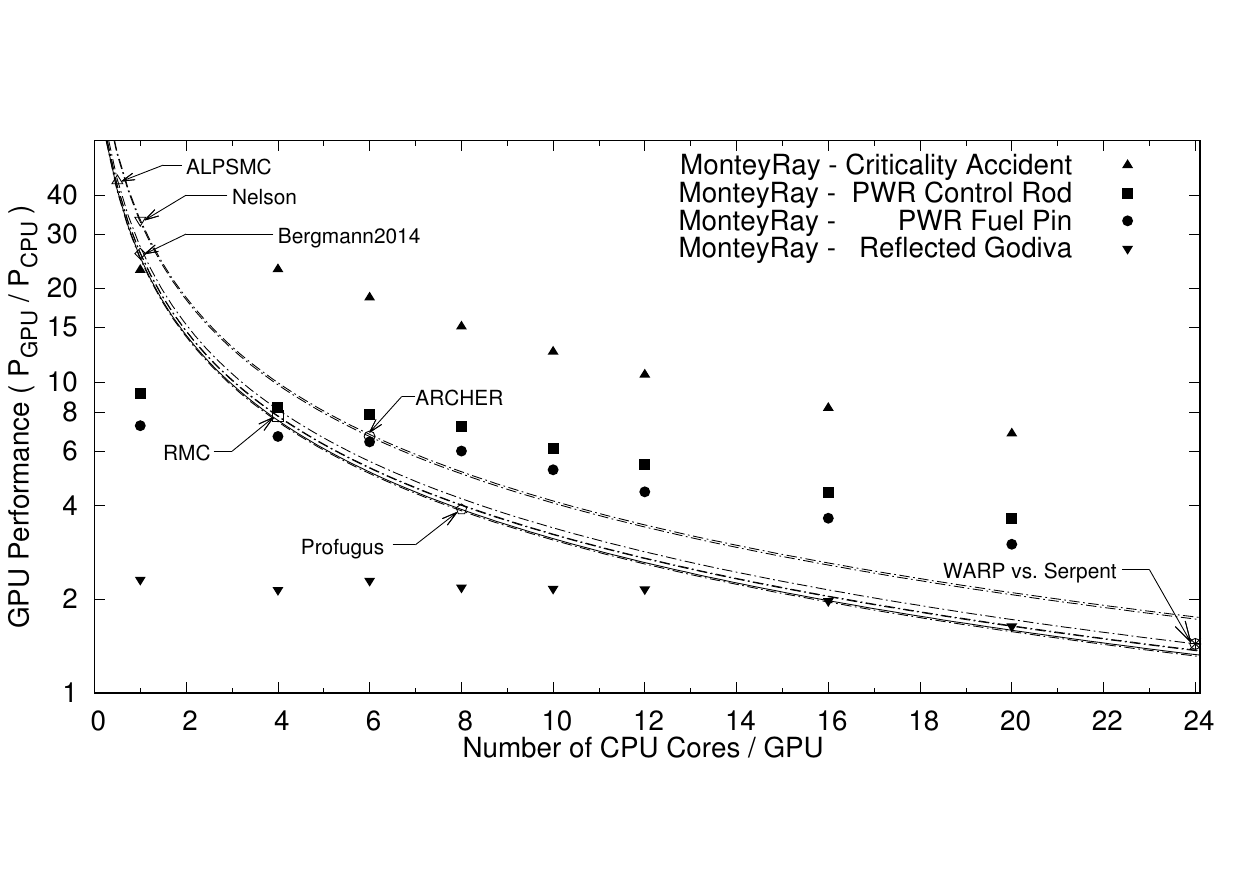}
  \caption{Results of MonteRay performance ($\blacktriangle, \blacktriangledown, \blacksquare, \bullet$  )  compared with other researchers implementations of neutron transport on GPUs.
  The performance of the other researchers' work, except Profugus, has been projected with dotted lines (\protect\dashedrule) for multiple CPU cores for using Eq.~\eqref{eq:gpu_ncpus}\@ and a strong scaling efficiency of 78\%.  
  The Profugus result have been projected with a solid line (\solidrule) and as it was the most full featured code that had direct CPU performance comparisons.}
  \label{fig:gpu_cpu_trend}
\end{figure*}

The results of ALPSMC, ARCHER, Bergmann, Nelson, Profugus, RMC, and ``WARP vs. Serpent'' have been plotted in \autoref{fig:gpu_cpu_trend} along with the MonteRay results as a function of the number of CPU cores per GPU\@.  
The performance of the other researchers' work has been projected for $n$ CPU cores for using Eq.~\eqref{eq:gpu_ncpus}.
The comparison with other researchers results indicate that MonteRay performs very well.   
The performance is exceptional given the fact that the goal of the \VRCE is to minimize the cost of accelerating production Monte Carlo codes with GPU hardware instead of maximizing GPU performance.
\textit{A word of caution:}  the MonteRay results should only be viewed in the context of acceleration of MCATK\@.   
If MonteRay were to be used with another Monte Carlo code the results would be different.    
These results should not be used to suggest that the \VRCE performs better than porting the random walk to the GPU for all Monte Carlo codes.
Legacy Monte Carlo codes and full featured production codes will benefit the most from using the \VRCE.
 
\subsection{Possible Impacts}

There are several possible impacts of the \VRCE: improving performance of calculations on GPUs for certain applications,  advancing Monte Carlo transport methods, and advancing Monte Carlo for high performance computing (HPC).

\subsubsection{Impacts on Applications}

The \VRCE on GPU hardware has the potential to speedup and/or produce higher fidelity neutron fluence tallies for Monte Carlo calculations over a wide range of applications.
The simulation of the PWR fuel assembly, presented in \autoref{pwrsection}, showed that the \VRCE was able to obtain significantly lower statistical uncertainty in strong absorbers, like control rods and burnable poisons.  
The uncertainty in the control rods was comparable to the uncertainty in neighboring fuel pins.   
This is a departure from the behavior of the uncertainty of the \TLE, which is typically inversely proportional to the neutron fluence.
Reducing the uncertainty in strong absorbers could be important to nuclear fuel designers who need to accurately calculate the depletion of  burnable poisons.  

The \VRCE obtains its highest performance when transporting neutrons through optically thin material, such as air or vacuum. 
This can  be applied to the transport of neutrons through atmosphere at large distances, such as calculating the dose in urban environments due to improvised nuclear devices or designing detectors for nuclear forensics.     
This is also useful in radiation protection for calculating the dose to personnel in nuclear facilities.

\subsubsection{Impacts on Monte Carlo neutron transport codes}

The \VRCE was designed for GPU hardware, but it may also increase performance on CPU hardware.
The \VRCE, when performed on CPU hardware, performed worse than the \TLE for some applications.  
But for other applications, like transport through optically thin cells, the \VRCE on CPU hardware performed better than the \TLEPERIOD.   
Testing of the \VRCE on CPU hardware was performed with MCATK, without the use of MonteRay.  
After each collision, the ray cast was performed instead of banking the rays for further processing.    
Banking the rays may improve the cache performance on CPU hardware and lead to increases in performance.   
One can imagine having some cores, or nodes, designated to perform only the \VRCE as the random walk is performed on different cores.  
This could lead to a more task based parallelism approach to Monte Carlo transport.   
Such an approach, but using the \TLE, has been investigated by \cite{RomanoTallyServers}.

As the \VRCE is an expected-value estimator, it has the significant advantage of being able to work with non-analog Monte Carlo variance reduction methods.   
One of the most commonly used variance reduction techniques for estimating global fluence is the weight-window method.
Weight-window values can be estimated from a previous Monte Carlo calculation or from the use of forward and adjoint deterministic transport methods~\citep{boothWeightWindows,cooperAndLarson,wagner2007forward}.  
Weight-windows and the \VRCE will be complementary techniques.  
Weight-windows could be used to enhance the \VRCE, using the weight window to vary the number of ray samples per collision based on particle location or energy.
The forced collision method could also be important for increasing the performance of the \VRCEPERIOD.  
Forced collisions can be used in regions of high importance but with low probabilities of interaction. 
In these cases the \VRCE would ensure that the forced collision makes contributions to many surrounding cells. 

The \VRCE has not yet been tested for the estimation of \keff.  
It is expected that using the \VRCE will result in lower variance than the \TLE of \keff.  
Without considering any impact on the estimate of \keff, the \VRCE will improve performance of eigenvalue calculations which also have global fluence or reaction rate tallies. 

\subsubsection{Impacts on High Performance Computing}

No general purpose production Monte Carlo neutron transport code has yet been ported to high performance computing (HPC) machines with accelerators.
The list of such general purpose production codes include the world class codes \MCNP~\citep{mcnp6}, KENO~\citep{keno}, MVP~\citep{mvp2013}, MONK~\citep{monk2013}, MCU~\citep{mcu}, and TRIPOLI~\citep{tripoli}\@.
These codes have been in development for 27-43 years now (or more) so their complexity makes them difficult to port to GPUs\@.
None of the more recent efforts at developing more modern Monte Carlo codes for HPC also support accelerators.  
These new codes include MCATK~\citep{mcatk}, Mercury~\citep{mercury}, MC21~\citep{mc21}, Shift~\citep{shift}, OpenMC~\citep{openmc2013}, and Serpent~\citep{serpent}. 
WARP is the only Monte Carlo neutron transport code designed for GPUs that can be classified as having full neutron physics using continuous energy cross sections.  
But WARP is still a specialty built research code.  

Conversely, deterministic neutron transport codes have been some of the first applications ported to leadership class HPC machines.  
Los Alamos National Laboratory ported its PARTISN deterministic transport code to the Cell Broadband Engine accelerators when it obtained Roadrunner in 2008, the first machine in the world to exceed one petaflops/s \citep{baker2017}.
And Oak Ridge National Laboratory ported its DENOVO deterministic transport code to use GPUs when Titan debuted on the Top500 list at \#1 in 2012~\citep{denovotitan}.  
However, the radical change in machine architecture signaled by the Roadrunner machine, did prompt the start of development of MCATK at Los Alamos in 2008 and has thus lead to the development and prototyping of the \VRCEPERIOD.    

Monte Carlo particle transport codes have not been targeted as science applications on leadership class HPC machines because it has been considered too expensive to adapt production Monte Carlo codes to accelerators~\citep{brown2011recent}.   
This may be true if one follows the conventional wisdom of porting the Monte Carlo random walk to the accelerator.    
However, a method like the \VRCE can significantly improve the performance when using an accelerator with only a few thousand lines of additional coding. 
The development of a full feature library that can be used as a plug-in package by multiple Monte Carlo codes would be on the order of one to a few person-years.
Such an investment is small considering the hundreds of person-years needed to develop a world class Monte Carlo code.  
This methodology for using accelerators also allows the production Monte Carlo code to maintain its significant investment in verification and validation (V\&V)\@.  
The V\&V performed over a 40 year period by a production code develop team and their tens of thousands of users should not be tossed aside merely because a new computer architecture is available.  

The second notion this work dispels is that Monte Carlo neutron transport requires double precision math for all calculations.   
In fact this work shows that the ray casting, which is the largest computational cost for the \VRCE, can be performed in single precision.   
Most of the tens of thousands of world wide users of Monte Carlo codes perform their calculations on desktop computers.   
Their desktop computers generally have consumer grade GPUs which excel at single precision operations.
HPC machines that use GPU accelerators use GPUs which excel at double precision operations like Nvidia's Tesla line of GPUs\@. 
The \VRCE would allow the tens of thousands of world wide users of Monte Carlo codes using desktop machines with consumer grade GPUs to see significant performance benefits.  
Consumer grade GPUs also provide a lower cost alternative to developing HPC machines specifically for Monte Carlo neutron transport.  
For example, MonteRay performed better on the lower cost Nvidia Titan X than on an Nvidia Tesla K40\@.  
~\cite{bergmann2017} also reported that WARP performed as well on the Nvidia Titan Black as on the Nvidia Tesla K80\@.   
 He pointed out that these consumer grade GPUs cost about  one third the price of the Tesla's and consume about  20\% less energy.  

\subsection{Future Research}

The next step in the evaluation of the \VRCE is to analyze its performance for the transport of photons.  
A number of groups have reported significant increase in performance on GPU hardware for the Monte Carlo transport of photons~\citep{gpumcd,henderson2013,ARCHER}\@. 
It is not expected that the \VRCE will be able to match these results, which can exceed 1000 in comparison to a single CPU core.
However, other researchers have questioned the fairness of typical GPU to CPU comparisons for photon transport as the codes are not often optimized for CPU hardware or codes specifically designed to GPUs are compared to more full featured codes running on CPU hardware~\citep{Jia2015}\@.  
Performance of the \VRCE for photons is expected to be better than for neutrons but will vary based on the optical thickness of cells over the photon energy spectrum.   
 
The offloading of ray casting to the GPU for next-event estimator estimators will also be demonstrated.
The next-event estimator is typically used for radiography simulations and for estimating fluence in areas that have few random walk particles.  
The next-event estimator~\citep{Kalos:63} requires a ray-cast from each source and collision event to the estimator location.   
It was this ray-cast and its expected performance on the GPU that was the inspiration for the \VRCEPERIOD.   
The next-event estimator places no additional burden on the CPU so the performance should be better than for the \VRCEPERIOD.  

The algorithms in MonteRay need to be optimized for the GPU\@.   
The \VRCE and ray casting algorithms have been directly taken from MCATK and ported to CUDA\@.
No attempt to optimize the coding has been made.
As the ray cast becomes faster, the performance on multiple CPU cores will approach the performance increase using a single CPU core paired with a GPU.
When the ray cast on the GPU is able to keep up with the random walk on all cores of a node, then performance of the \VRCE will be limited by the rate of sampling the rays from collision on the CPU\@.     

MonteRay only demonstrates the \VRCE as a proof of concept.   
Several improvements must be made to make it useful for production Monte Carlo codes.
MonteRay must support coupled neutron and photon transport.   
Its geometry must be extended.  
It must support 1-D spherical meshes, 2-D cylindrical (R-Z) meshes, and the solid body geometry of MCATK\@. 
It should also support \MCNP, which has structured mesh geometry, unstructured mesh geometry, and constructive solid geometry.   
MCNP's structured mesh capability is similar to MCATK's and should not prove difficult to port to the GPU\@. 
It is not feasible to consider porting MCNP's constructive solid geometry to CUDA, but it may be possible to include MCATK's solid body geometry as an optional geometry type within \MCNPPERIOD.
This should not be difficult as MCATK is designed as a toolkit to provide functionality to codes such as \MCNPPERIOD.
MCATK's solid body geometry is defined by a scene graph and would be easier to port to CUDA and other accelerator supported libraries~\citep{Trahan2015PHYSOR}\@.     

\section{Conclusions} 

The \VRCE demonstrates a  method of accelerating existing Monte Carlo particle transport codes with GPU hardware.  The \VRCE can be introduced into an existing Monte Carlo transport code as a library without significant changes to the Monte Carlo code base.   
This significantly reduces the cost of enabling Monte Carlo codes to utilize GPU hardware and maintains the V\&V history of the software.

Depending on the complexity of the Monte Carlo code, the \VRCE may perform better than porting the Monte Carlo random walk to GPU hardware.
A review of other research shows that porting the Monte Carlo random walk to the GPU can obtain a $\sim$4-5 times speedup on a single GPU as compared to eight CPU cores.
The \VRCE can obtain higher performance but the performance is dependent on the Monte Carlo code and the problem being simulated.
The estimator shows the best performance for simulations with optically thin cells, such as problems in the areas of radiation protection and atmospheric transport.   
It shows the worst performance for simulations with optically thick cells.   
For a criticality accident scenario, where the neutron fluence in air was calculated within a concrete room, a performance improvement of 23x was obtained for a single CPU core paired with a GPU (\autoref{tab:results_all_problems}).   For eight CPU cores / GPU the performance increase was 15x\@.
Simulation of a PWR fuel assembly showed performance increases of 9.2 and 7.2 for one and eight CPU cores respectively.
And a simulation of the Reflected Godiva benchmark showed increases of 2.3 and 2.2 for one and eight CPU cores.

%

\begin{table}[tbp]
\begin{center}
\begin{tabular}{x{1.0cm}x{1.5cm}x{1.5cm}x{1.4cm}x{1.4cm}}
\toprule
 & \multicolumn{4}{c}{VRC Estimator Performance ($\eta$)} \\
\cmidrule{2-5}
\begin{minipage}[b]{1.0cm}\setlength{\baselineskip}{0.8\baselineskip}  \centering Number of CPU cores per GPU  \end{minipage} &
\begin{minipage}[b]{1.5cm}\setlength{\baselineskip}{0.8\baselineskip}  \centering PWR Fuel Assembly - Fuel \\ Pin \\ Tally\end{minipage} &
\begin{minipage}[b]{1.5cm}\setlength{\baselineskip}{0.8\baselineskip}  \centering PWR Fuel Assembly - Control Rod Tally\end{minipage} &
\begin{minipage}[b]{1.4cm}\setlength{\baselineskip}{0.8\baselineskip}  \centering Criticality Accident Scenerio \end{minipage} &
\begin{minipage}[b]{1.4cm}\setlength{\baselineskip}{0.8\baselineskip}  \centering Reflected Godiva \end{minipage} \\
\midrule
  1 & 7.3 & 9.2 & 23.0 & 2.3 \\
  4 & 6.7 & 8.3 & 23.1 & 2.2 \\
  6 & 6.4 & 7.9 & 18.7 & 2.3 \\
  8 & 6.0 & 7.2 & 15.1 & 2.2 \\
10 & 5.2 & 6.1 & 12.5 & 2.2 \\
12 & 4.4 & 5.4 & 10.6 & 2.2 \\
16 & 3.7 & 4.4 &  8.3 & 2.0 \\
20 & 3.0 & 3.6 &  6.8 & 1.6 \\
\bottomrule
\end{tabular}
\end{center}
\caption{Measured performance of the \VRCE for the three simulations, a PWR fuel assembly with two tally regions, a criticality accident scenario, and the Reflected Godiva benchmark problem. }
\label{tab:results_all_problems}
\end{table}%

\section{Acknowledgments}
This work was demonstrated with LANL's Monte Carlo Application Toolkit (MCATK), developed by Steve Nolen, Terry Adams, 
Travis Trahan, Jeremy Sweezy, Lori Pritchett-Sheats, and Chris Werner.  
MCATK development has been funded by LANL's Advanced Simulation and Computing (ASC) program.  
GPU performance testing was performed using LANL's Darwin HPC cluster, an ASC R\&D test bed, managed by Ryan Braithwaite.

Funding for the evaluation of the \VRCE was provided by Los Alamos National Laboratory's Pathfinder funding program.
This material is based upon work supported by the U.S. Department of Energy, Office of Science, National Nuclear 
Security Administration, under contract number DE-AC52-06NA25396.

This document is authorized for unlimited release as Los Alamos Report Number LA-UR-17-22573.
\url{http://permalink.lanl.gov/object/view?what=info:lanl-repo/lareport/LA-UR-17-22573}
\bibliography{references}

\end{document}